\documentclass[aps,prd,amsmath,amssymb,superscriptaddress,nofootinbib,showpacs,preprintnumbers,notitlepage]{revtex4-1}
\usepackage{graphicx,hyperref,color,subfigure,soul}
\usepackage{cleveref,ctable,braket,xspace}
\usepackage{slashed}
\usepackage{appendix}

\usepackage{booktabs}
\usepackage{dcolumn}

\AtBeginDocument{
  \heavyrulewidth=.08em
  \lightrulewidth=.05em
  \cmidrulewidth=.03em
  \belowrulesep=.65ex
  \belowbottomsep=0pt
  \aboverulesep=.4ex
  \abovetopsep=0pt
  \cmidrulesep=\doublerulesep
  \cmidrulekern=.5em
  \defaultaddspace= .5em
  \setlength{\tabcolsep}{0.5em}
}

\hypersetup{ 
    pdfnewwindow=true,      
    colorlinks=true,       
    linkcolor=blue,          
    citecolor=blue,        
    filecolor=blue,      
    urlcolor=blue        
}  

\crefname{figure}{Fig.}{Figs.}
\crefname{equation}{Eq.}{Eqs.}
\crefname{table}{Table}{Tables}
\crefname{chapter}{Chapter}{Chapters}
\crefname{section}{Section}{Sections}
\crefname{appendix}{Appendix}{Appendices}

\newcommand{\nn}{\nonumber}

\newcommand{\valencia}{Departamento de F\'\i sica Te\'orica and IFIC, 
Centro Mixto Universidad de Valencia-CSIC, 
Institutos de Investigaci\'on de Paterna, E-46071 Valencia, Spain}
\newcommand{\mainz}{Institut f\"ur Kernphysik \& PRISMA Cluster of Excellence,
Johannes Gutenberg Universit\"at, D-55099 Mainz, Germany}

\newcolumntype{d}{D{.}{.}{3.8}}
\newcommand{\none}{\multicolumn{1}{c}{---}}

\allowdisplaybreaks

\begin{document}

\title{Systematic study of octet-baryon electromagnetic form factors
in covariant chiral perturbation theory}

\author{A.~N.~\surname{Hiller Blin}}
\affiliation{\valencia}
\affiliation{\mainz}
	
\begin{abstract}
We perform a complete and systematic calculation of the octet-baryon form factors within the fully covariant approach of $SU(3)$ chiral perturbation theory at $\mathcal{O}(p^3)$. We use the extended on-mass shell renormalization scheme, and include explicitly the vector mesons and the spin-3/2 decuplet intermediate states. Comparing these predictions with data including magnetic moments, charges, and magnetic radii, we determine the unknown low-energy constants, and give predictions for yet unmeasured observables, such as the magnetic moment of the $\Sigma^0$, and the charge and magnetic radii of the hyperons.
\end{abstract}
\date{\today}
\maketitle

\section{Introduction}

The electromagnetic structure of hadrons plays a fundamental role in our understanding of the structure of matter and the underlying strong interaction. At high energies, the fundamental properties of hadrons are well described by perturbative quantum chromodynamics (pQCD). However, in order to describe interactions at distances larger than the size of hadrons, corresponding to energies well below 1~GeV, pQCD breaks down, and suitable low-energy effective-field theories (EFTs) are useful. This energy regime is the focus of this paper, and we use the methods of fully covariant $SU(3)$ baryon chiral perturbation theory (ChPT) for our description of the hadron properties~\cite{Weinberg:1978kz,Gasser:1987rb,Krause:1990xc}. We use this EFT of QCD including explicitly the decuplet of spin-3/2 baryons~\cite{Jenkins:1990jv,Pascalutsa:2000kd,Pascalutsa:2002pi,Pascalutsa:2006up} and the vector-meson contributions~\cite{Borasoy:1995ds,Drechsel:1998hk,Kubis:2000aa,Kubis:2000zd,Schindler:2005ke,Bauer:2012pv}.

At each order in the chiral expansion, the interactions are parameterized by a set of low-energy constants (LECs), many of which have already been determined from data. When including baryons in ChPT, the chiral power counting in terms of momenta and masses seems to be spoiled~\cite{Gasser:1987rb}. This is due to the baryon masses being of the order of 1~GeV. As a result, a priori there is no clear way to associate a specific chiral order with a definite number of loops. Several renormalization schemes to solve this issue have been proposed in the past. The first such approach is semi-relativistic, which exploits that the baryons are much heavier than the Goldstone bosons, and hence an expansion in the inverse baryon mass is made. This is called heavy-baryon ChPT (HBChPT), first introduced in Ref.~\cite{Jenkins:1990jv}. Covariant approaches are slightly more involved, and resum terms of higher order from the HBChPT viewpoint. The infrared regularization (IR) scheme separates the loops into infrared and regular parts, obtained by a manipulation of the baryon propagators~\cite{Becher:1999he,Becher:2001hv,Scherer:2002tk,Ando:2006xy}. The regular parts fully encode the power-counting breaking terms (PCBTs), and are thus absorbed in the LECs of the most general Lagrangians. This approach is based on the works in Refs.~\cite{Tang:1996ca,Ellis:1997kc}. However, we choose the extended on-mass shell (EOMS) scheme~\cite{Gegelia:1999gf,Fuchs:2003qc}, which is known to converge well for a range of processes~\cite{Fuchs:2003ir,Lehnhart:2004vi,Schindler:2006it,Schindler:2006ha,Geng:2008mf,Geng:2009ik,MartinCamalich:2010fp,Alarcon:2011zs,
Ledwig:2011cx,Chen:2012nx,Alvarez-Ruso:2013fza,Ledwig:2014rfa,Lensky:2014dda,Blin:2014rpa,Blin:2016itn,HillerBlin:2016jpb}. This scheme relies on the knowledge that the PCBTs that spoil the chiral series have fully analytical expressions. Therefore, they can be identified with terms of the Lagrangian, and absorbed into the corresponding LECs by an extension of the minimal subtraction (MS) scheme of dimensional regularization.

Some of the differences between the EOMS and IR schemes are compensated by adjustments in the values of the LECs, see also the discussions in Refs.~\cite{Ando:2006xy,Ledwig:2011cx,Schindler:2003xv,Scherer:2012xha,Siemens:2016jwj}. The renormalization of the divergences is treated in the same way, the difference lies in which subset of the finite terms is absorbed into the LECs. The EOMS scheme absorbs only the minimal amount of terms needed to restore the power counting, thus maintaining the analytical structure of the amplitudes. This is not the case for the IR scheme, which is often used as an argument against its usage.

The baryon electromagnetic form factors are obtained from the reaction amplitude which describes a virtual photon coupling to these hadrons. At small momentum transfer squared $q^2$, where a Taylor expansion of the amplitude in terms of $q^2$ is reliable, the coefficients of this expansion yield insight about quantities such as the charge radii and the electromagnetic moments. For non-vanishing photon virtualities, one can relate them to the charge and magnetic densities. A good general review on nucleon electromagnetic form factors is given in Refs.~\cite{Perdrisat:2006hj,Punjabi:2015bba}, both from the theoretical and from the experimental points of view. Assuming $CP$ conservation and on-shell baryons, the electromagnetic current of spin-1/2 baryons is determined by only the Dirac and the Pauli form factors, or combinations thereof.

The first measurements of the nucleon form factors are described in Refs.~\cite{Hofstadter:1955ae,Hofstadter:1956qs,Yearian:1958shh}. In Ref.~\cite{Hofstadter:1956qs}, it was shown how to extract the form factors from the ratio between the experimental cross section and the expected Mott cross section of pointlike particles. The separate electric and magnetic form factors were obtained for the first time by the intersecting ellipse method, described by Hofstadter in 1955~\cite{Hofstadter:1959nya}. Since then, countless efforts have been made to extract form factors experimentally, at facilities such as the Stanford Linear Accelerator Center (SLAC), the Deutsches Elektronen-Synchrotron (DESY), the Nationaal Instituut voor Kernfysica en Hoge Energie Fysica (NIKHEF), and the Jefferson Lab (JLab), among many others. More recently, high-precision experiments with electron beams have been performed, e.g.\ at the Mainz Microtron (MAMI).

The nucleon form factors have gathered considerable attention lately due to apparently conflicting results. Measurements of the proton charge radius $r_E^p$ via electron-scattering experiments~\cite{Bernauer:2010wm} have shown a disagreement with the results from precise atomic measurements of the muonic hydrogen Lamb shift~\cite{Pohl:2010zza}. This is commonly known as the proton radius puzzle. The first method is based on writing the electric form factor $G_E$ as a function of the squared momentum transfer $q^2$. The charge radius is then obtained from the slope of $G_E$ at vanishing $q^2$. The second method uses lasers to induce atomic transitions, where the charge radius is related to the size of the gap between the levels. In fact, the mean value reported from electron-scattering experiments is of $r_E^p=0.8751(61)$~fm~\cite{Bernauer:2010wm}, while the atomic measurement yielded $r_E^p=0.84184(67)$~fm~\cite{Pohl:2010zza}. Interestingly, similar measurements on normal instead of muonic hydrogen were compatible with the electron-scattering result. The main difference between the two types of hydrogen is that, since the muon mass is approximately 200 times larger than the electron's, the Bohr radius is 2 orders of magnitude smaller for the muonic hydrogen. Thus, the muon is much more sensitive to the proton radius, yielding a higher empirical precision. But even when taking into account the larger uncertainty of the electron-scattering measurements, the discrepancy between results is of the order of $5\,\sigma$. This has led to speculations about possible new physics explanations, for instance through dark matter coupling to muons~\cite{Jentschura:2010ha}.

The atomic method has been reanalysed in Refs.~\cite{Jentschura:2010ha,Distler:2010zq,Carlson:2011zd,Miller:2011yw,Alarcon:2013cba}, while there has also been an effort to better understand the results from electron scattering~\cite{Griffioen:2015hta}. For the latter, an experimental determination of the slope of $G_E$ at the exact point of real photons $q^2=0$ is of course not possible. However, one can gather probes of very small virtualities, and extrapolate the value to the physical point. It goes without saying that this extrapolation leads to some model dependence. One could argue that, at least when leaving the region where $q^2$ is nearly zero, one needs to take into account the singularities that appear in the complex plane. The first such point is the two-pion production threshold, where the virtual photon couples to the baryon via two pions. Therefore, a polynomial fit for the extrapolation would only be reliable for momentum transfers significantly lower than this cut, where data are scarce. Nevertheless, in Ref.~\cite{Griffioen:2015hta} such an analysis has been done, leading to a result compatible with those from the muonic hydrogen Lamb shift, thus displaying a possible solution for the issue described above.

In order to be able to reproduce the behaviour observed at higher photon virtualities, one needs to find a theoretical approach which describes the complex-plane singularities as well. This question has been studied within dispersion analyses~\cite{Mergell:1995bf,Hammer:2003ai,Belushkin:2005ds,Belushkin:2006qa}, quark models~\cite{Silva:2005vp,Ramalho:2011pp,Liu:2013fda} and with lattice QCD~\cite{Yamazaki:2009zq,Syritsyn:2009mx,Alexandrou:2011db,Collins:2011mk,Bratt:2010jn}, among other approaches~\cite{Flores-Mendieta:2015wir,Carrillo-Serrano:2016igi,Aliev:2013jda}. In the present work we use the methods of ChPT to answer the same question. This effective description of pion loops ensures that the two-pion cut is taken into account in a consistent manner~\cite{Krause:1990xc,Kaiser:2003qp,Fuchs:2003ir,Bernard:1998gv,Jiang:2009jn,Kubis:2000zd,Ledwig:2011cx}. When moving to yet higher virtualities, kaon loops~\cite{Kubis:1999xb,Geng:2008mf,Geng:2009hh} or even vector-meson exchanges should be considered as well~\cite{Kubis:2000aa,Kubis:2000zd,Schindler:2005ke,Bauer:2012pv}.

We focus on the extraction of the electromagnetic form factors of all the octet baryons, in order to obtain insight into the inner charge and magnetic distributions of these hadrons. This is particularly interesting because recent progress in lattice calculations allows for an independent calculation and comparison of the results~\cite{Shanahan:2014uka,Shanahan:2014cga}. Furthermore, the work presented here may have an impact on calculations that use form factors in order to obtain information about more complicated processes, such as Compton scattering, see Ref.~\cite{Gasser:2015dwa}. There, the results of the form factors are used in order to determine the structure functions needed to obtain the nucleon polarizabilities in a dispersive representation. Other interesting observables to study are the baryon charge and magnetic densities. With dispersion theory, one can relate them to the imaginary parts of the form factors. The calculation of the peripheral transverse densities has been performed both in $SU(2)$ and in $SU(3)$~\cite{Strikman:2010pu,Granados:2013moa,Alarcon:2017lkk,Alarcon:2017asr,HillerBlin:2016jpb,Granados:2017cib}.

In a similar approach to ours, in Ref.~\cite{Geng:2009hh} the magnetic moments of the octet baryons were calculated, most of which have also already been experimentally extracted. Thus, one can constrain the LECs very well, and test the compatibility of the ChPT approach with the data. We extend the authors' to further observables, the charge and magnetic radii, for which we additionally include the effects of the vector mesons, in a model-dependent approach: the values of the couplings of the vector mesons to the baryons depend on the phenomenological method with which they are extracted. We chose those values obtained in Ref.~\cite{Chiang:2001as}, but also studied the effect on the results when choosing e.g.\ the Bonn-potential values~\cite{Machleidt:1987hj,Machleidt:2000ge}. In contrast, in Ref.~\cite{Bauer:2012pv} it has been discussed how to include these spin-1 fields in a self-consistent manner. While the vector mesons do not contribute to the magnetic moments, in the case of the radii they turn out to give important corrections.

This paper is structured as follows: in \cref{Sform}, we introduce the formalism for the calculation of the baryon form factors and the related observables. More precisely, in \cref{SformSamp}, we show the model-independent decomposition of the electron-scattering amplitude into the form factors, and its connection to the observables studied, such as the charge radii. We discuss the ChPT framework in more detail in \cref{SformSchpt}. In \cref{SformSvec}, we introduce the formalism needed for the vector-meson contributions. The results are reviewed in \cref{Sres}. Namely, we give the results for the electric form factors and related observables, and compare them to those found experimentally and in other theoretical works. The explicit expressions for the form factors are given in App.~\ref{Sappamp}. In \cref{Ssum}, we present a summary and outlook of this work.

\section{Formalism}\label{Sform}
\subsection{Amplitude decomposition}\label{SformSamp}
The electromagnetic form factors of the octet baryons are given by the Lorentz-invariant decomposition of the matrix element of the vector current $J^\mu$ between baryon states, which reads:\footnote{There is an additional structure appearing, proportional to $q^\mu$, see also Refs.~\cite{Bincer:1959tz,Krause:1990xc,Koch:2001ii,Haberzettl:2011zr}. This term is necessary to fulfill the Ward-Takahashi identities, but can be effectively dropped in most cases, e.g.\ when the baryons are on shell.}
\begin{align}\label{Eq:EM-current-matrix-element}
\bra{B(p')}J^\mu\ket{B(p)}=\bar{u}(p')\left(\gamma^\mu F_1(q^2)+\frac{\mathrm{i}\sigma^{\mu\nu}q_\nu}{2m_{B0}} F_2(q^2)\right)u(p),
\end{align}
where $u(p)$ is the spinor of the octet baryon with mass $m_{B0}$, and $F_1$ and $F_2$ its electromagnetic form factors. The photon momentum is given by $q=p'-p$, where $p'$ and $p$ are the outgoing and incoming baryon momenta, respectively. The Mandelstam variable $t$ is given by the momentum transfer squared $q^2$. As usual, $\sigma^{\mu\nu}=\frac{\mathrm{i}}{2}\left[\gamma^\mu,\gamma^\nu\right]$.

The baryon electric form factor is then given by
\begin{equation}
G_E(q^2)=F_1(q^2)+\frac{q^2}{4m_{B0}^2}F_2(q^2)=c_b+q^2\frac{\left<r_E^2\right>}{6}+\frac{q^4}{2}\frac{\mathrm{d}^2}{(\mathrm{d}q^2)^2}G_E(q^2)|_{q^2=0}+\mathcal{O}(q^6),\label{EqDDG0}
\end{equation}
where $c_b$ and $r_E$ are the baryon charge and charge radius, and the pseudoscalar meson-loop contribution to $\frac{\mathrm{d}^2}{(\mathrm{d}q^2)^2}G_E(q^2)$ can be given as a prediction of ChPT, see Ref.~\cite{HillerBlin:2016jpb}. Therefore, the inclusion of this analytical function in the extrapolation of electron-scattering data can be of relevance for obtaining the proton charge radius. Experimental estimates for the value of the proton $\frac{\mathrm{d}^2}{(\mathrm{d}q^2)^2}G_E(q^2)|_{q^2=0}$ were given, for instance in Refs.~\cite{Higinbotham:2015rja,Bernauer:2010zga}. As for the baryon magnetic form factor, it is defined as
\begin{equation}
G_M(q^2)=F_1(q^2)+F_2(q^2).
\end{equation}

The observables we will be focusing on in this work are the baryon magnetic moment $\mu^B=G_M^B(q^2=0)$ in units of the nuclear magneton, and the charge and magnetic radii defined as
\begin{align}
\left<r_E^2\right>&=\left.\frac{6}{G_E(0)}\frac{\mathrm{d}G_E(q^2)}{\mathrm{d}q^2}\right|_{q^2=0},\quad\quad
\left<r_M^2\right>=\left.\frac{6}{G_M(0)}\frac{\mathrm{d}G_M(q^2)}{\mathrm{d}q^2}\right|_{q^2=0}
\intertext{
for $G_E(0)\neq 0$, and as}
\left<r_E^2\right>&=\left.6\frac{\mathrm{d}G_E(q^2)}{\mathrm{d}q^2}\right|_{q^2=0}
\end{align} 
for $G_E(0)= 0$.

\subsection{Chiral perturbation theory}\label{SformSchpt}
In $SU(3)$, the Lagrangians involve the pseudoscalar octet mesons $\phi$, the octet baryons $B$, the decuplet baryons $T_\mu$
and the photon fields $v_\mu=-e\mathcal{A}_\mu Q$ for $e>0$. The explicit forms of the corresponding matrices in terms of the Gell-Mann matrices can be found in the literature, e.g.\ in Refs.~\cite{Ledwig:2014rfa,HillerBlin:2016jpb}. The lowest-order chiral Lagrangian involving 
only photons and the two hadron octets reads 
\begin{equation} \label{Lagr_full} 
  \mathcal{L} = \mathcal{L}^{(2)}_{\phi\phi} + \mathcal{L}^{(1)}_{\phi B},
\end{equation}
where 
\begin{equation}
  \mathcal{L}^{(2)}_{\phi\phi}
  =\frac{F_0^2}4\text{Tr}\left(u_\mu u^\mu +\chi_+\right)
  \label{eqmesonlag}
\end{equation} 
is the $\mathcal O(p^2)$ meson Lagrangian, and 

\begin{equation}
  \mathcal{L}^{(1)}_{\phi B}
  =\text{Tr}\left(\bar{B}(\mathrm{i}\slashed{\mathrm{D}}-m_{B0})B\right)+
  \frac D2\text{Tr}\left(\bar{B}\gamma^\mu\gamma_5\left\{u_\mu,B\right\}\right)+
  \frac F2\text{Tr}\left(\bar{B}\gamma^\mu\gamma_5\left[u_\mu,B\right]\right)
  \label{eqlagmesbar}
\end{equation}
is the $\mathcal O(p^1)$ Lagrangian that includes octet baryons. The commutator and anticommutator refer to flavour space. Here, $m_{B0}$ and $F_0$ denote the baryon-octet mass and the meson-decay constant, respectively, both in the chiral limit. The vielbein $u_\mu$ and the covariant derivative $\mathrm{D}_\mu$ read
\begin{align}
 u_\mu&=\mathrm{i}\left\{u^\dagger,\nabla_{\mu} u\right\},\quad
\mathrm{D}_\mu B = \partial_\mu B + [\Gamma_\mu,B],
\intertext{where}
 \nabla_{\mu} u &= \partial_\mu u - \mathrm{i}(v_\mu + a_\mu)u 
+ \mathrm{i}u(v_\mu - a_\mu),\quad
 \Gamma_\mu = \frac12[u^\dagger,\partial_\mu u] 
- \frac{\mathrm{i}}{2}u^\dagger(v_\mu + a_\mu)u 
- \frac{\mathrm{i}}{2}u(v_\mu - a_\mu)u^\dagger.\label{Eq:chirchon}
\end{align}
When working exclusively with external photon fields, the axial field $a_\mu$ can be set to zero.
The LECs $D$ and $F$ are determined from nucleon and hyperon 
$\beta$ decays, where the combination $F+D$ corresponds to the LEC $g_0$ in the $SU(2)$ limit. 

In this article, the terms needed from higher-order Lagrangians are those that couple the photon to the baryon directly. They have been constructed in Refs.~\cite{Krause:1990xc,Kubis:2000aa,Oller:2006yh,Frink:2006hx,Oller:2007qd}, and read
\begin{align}
\mathcal{L}^{(2)}_{\gamma B}=&\frac{b_F}{8m_{B0}}\left<\bar{B}\left[f_+^{\mu\nu},\sigma_{\mu\nu}B\right]\right>+\frac{b_D}{8m_{B0}}\left<\bar{B}\left\{f_+^{\mu\nu},\sigma_{\mu\nu}B\right\}\right>,\\[1em]
\mathcal{L}^{(3)}_{\gamma B}=&\mathrm{i}\frac{d_{101}}{2m_{B0}}\left<\bar{B}\left[\left[\mathrm{D}_\mu, f_+^{\mu\nu}\right],\left[\mathrm{D}_\nu, B\right]\right]\right>+\mathrm{i}\frac{d_{102}}{2m_{B0}}\left<\bar{B}\left\{\left[\mathrm{D}_\mu, f_+^{\mu\nu}\right],\left[\mathrm{D}_\nu, B\right]\right\}\right>+\mathrm{H.c.},
\intertext{where}
\nn f_+^{\mu\nu}=&uF_L^{\mu\nu}u^\dagger+u^\dagger F_R^{\mu\nu}u,\\[1em]
\nn F_R^{\mu\nu}=&\partial^\mu \left(v^\nu+a^\nu\right)-\partial^\nu \left(v^\mu+a^\mu\right)-\mathrm{i}\left[\left(v^\mu+a^\mu\right),\left(v^\nu+a^\nu\right)\right],\\[1em]
 F_L^{\mu\nu}=&\partial^\mu \left(v^\nu-a^\nu\right)-\partial^\nu \left(v^\mu-a^\mu\right)-\mathrm{i}\left[\left(v^\mu-a^\mu\right),\left(v^\nu-a^\nu\right)\right],
\end{align}
and $b_D$, $b_F$, $d_{101}$, and $d_{102}$ are low-energy constants. Note that the third-order piece, while being consistent in all literature up to order $\mathcal{O}(p^3)$, has two different representations that give distinct higher-order contributions. Here we choose to use the Lagrangian in Ref.~\cite{Kubis:2000aa}, which reproduces the results commonly obtained in other works in $SU(2)$.\footnote{This means that there are contributions to both the form factors $F_1$ and $F_2$ from this third-order Lagrangian. If one were to use the other Lagrangian~\cite{Oller:2006yh,Frink:2006hx,Oller:2007qd}, only one of the two structures would survive, $F_1$, while the other would vanish due to being a fourth-order correction.}

In the present paper, the baryon decuplet is also included. 
The relevant terms of the Lagrangian that couples these decuplet 
fields $T_\mu$ to the octets of baryons and mesons are given 
in Refs.~\cite{Geng:2009hh,Geng:2009ys,Ledwig:2014rfa}, where the lowest-order terms needed read
\begin{align}
\mathcal{L}_{T\phi}^{(1)}=&\bar{T}_\mu^{abc}(\mathrm{i}\gamma^{\mu\nu\alpha}\mathrm{D}_\alpha-M_\Delta\gamma^{\mu\nu})T_\nu^{abc},
\\[1em]
\mathcal{L}_{TB\phi}^{(1)}=&\frac{\mathrm{i}\mathcal{C}}{M_\Delta}\epsilon^{ilm}\left[(\partial_\mu\bar{T}_\nu^{ijk})\gamma^{\mu\nu\rho}u_\rho^{jl}B^{km}+\mathrm{H.c.}\right].
\end{align}
Here,
\begin{align}
\gamma^{\mu\nu}=\frac{1}{2}\left[\gamma^{\mu},\gamma^{\nu}\right],\quad
\gamma^{\mu\nu\rho}=\frac{1}{2}\left\{\gamma^{\mu\nu},\gamma^{\rho}\right\},\quad
\gamma^{\mu\nu\rho\sigma}=\frac{1}{2}\left[\gamma^{\mu\nu\rho},\gamma^{\sigma}\right].
\end{align}
The covariant derivative acts on the decuplet as
\begin{align}
\nn\mathrm{D}_\alpha T_\nu^{abc}&=\partial_\alpha T_\nu^{abc} + (\Gamma_\alpha,T_\nu)^{abc},\\[1em]
(X,Y)^{abc}&=X^{ad}Y^{dbc}+X^{bd}Y^{adc}+X^{cd}Y^{abd}.
\end{align}

When performing the calculations with the $SU(3)$ Lagrangian, and then setting the kaon and $\eta$ loops to zero, one reproduces the $SU(2)$ result with the LEC correspondence $D+F=g_A$ and $\mathcal{C}=-\frac{h_A}{2\sqrt{2}}$. Nevertheless, when including the additional $SU(3)$ loops, a new fit to decay-width data has to be performed~\cite{Ledwig:2014rfa}, and those new values should be used for $F$, $D$ and $\mathcal{C}$ in the calculations.

When computing loops that include internal baryon lines, PCBTs might arise. This is referred to as the baryon ChPT power-counting problem~\cite{Gasser:1987rb}. A diagram of nominal order $N$ might after integration contain terms of order $n<N$. These terms spoil the convergence of the chiral series, and therefore must be identified and renormalized. Here, this is done in the extended on-mass shell (EOMS) scheme where, together with the divergences, these analytical expressions are absorbed into the LECs of the lower-order Lagrangians. The identification of these terms is best done by expanding the result as a series in small external momenta and masses, and then isolating the terms of order $n<N$.

Special care is needed when taking the spin-3/2 states into account. Besides the pion mass and the external momenta, another small parameter appears, $\delta= M_\Delta - m_{B0}\approx 300$~MeV, which is heavier than $m_\pi\approx 140$~MeV, but small when compared to the spontaneous symmetry-breaking scale $\Lambda\sim m_{B0}$. The propagator for a spin-$3/2$ state with four-momentum $p^\mu$ takes the Rarita-Schwinger form, see e.g.\ Ref.~\cite{Ledwig:2011cx}, and is of the order $\delta^{-1}$. In the range of external energies considered, it is reasonable to treat $\delta$ as being of the same order as those energies, $\mathcal{O}(p)$. This approach is called small scale expansion~\cite{Hemmert:1996xg,Hemmert:1997ye}. The power $N$ of a diagram with $L$ loops, $V^{k}$ vertices from a Lagrangian $\mathcal{L}^{(k)}$ of order $k$, $N_\pi$ mesonic propagators, $N_N$ octet-baryon propagators and $N_\Delta$ propagators for the decuplet is therefore counted as
\begin{align}
N=4L+\sum_{k=1}^{\infty}{kV^{k}}-2N_\pi-N_N-N_\Delta.\label{EOrderEps}
\end{align}
%
%
%
%
%

\begin{center}
\begin{figure}
\includegraphics[width=0.2\textwidth]{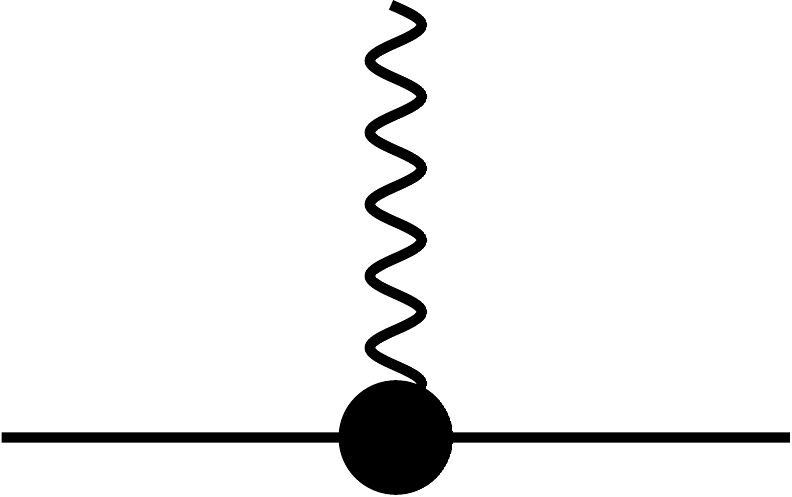}
\caption{Tree-level diagram that contributes to the baryon form factors. The vertex runs through orders $\mathcal{O}(p^1)$ to $\mathcal{O}(p^3)$. The wavy and continuous lines correspond to photon and octet-baryon fields, respectively.}
\label{Fdiagschpttree}
\end{figure}
\end{center}

\begin{center}
\begin{figure}
\subfigure[]{\label{Fdiagschpta}
\includegraphics[width=0.2\textwidth]{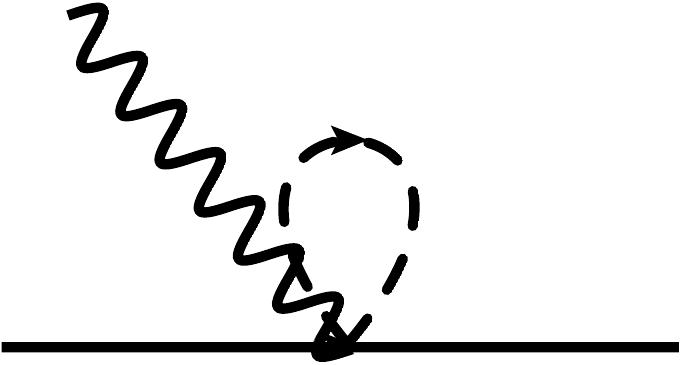}}
\subfigure[]{\label{Fdiagschptb}
\includegraphics[width=0.2\textwidth]{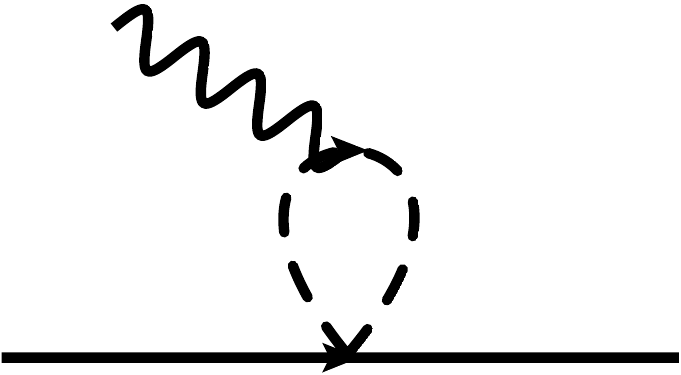}}
\subfigure[]{\label{Fdiagschptc}
\includegraphics[width=0.2\textwidth]{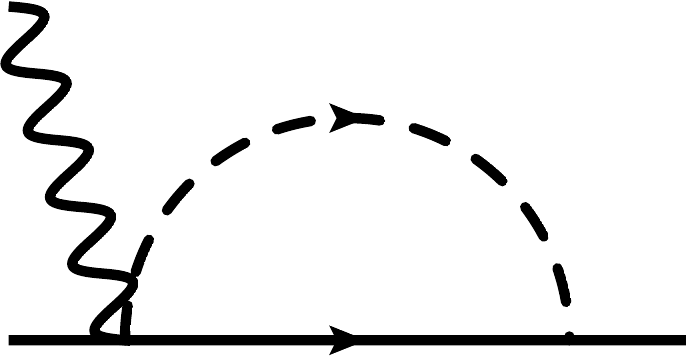}}
\subfigure[]{\label{Fdiagschptd}
\includegraphics[width=0.2\textwidth]{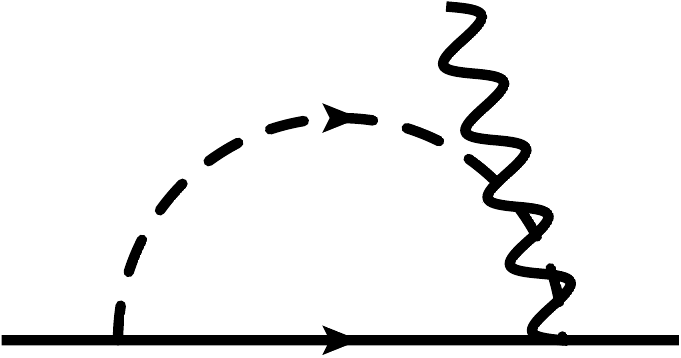}}
\subfigure[]{\label{Fdiagschpte}
\includegraphics[width=0.2\textwidth]{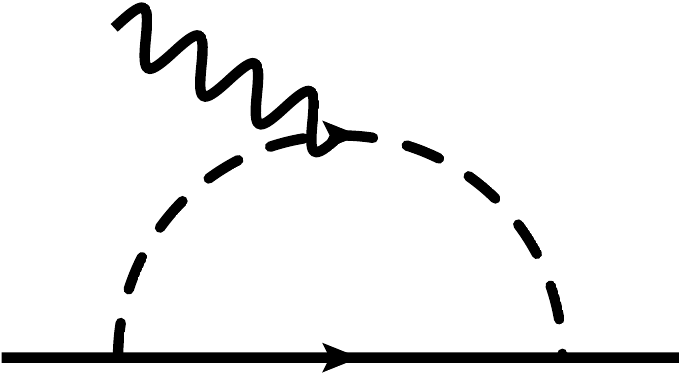}}
\subfigure[]{\label{Fdiagschptf}
\includegraphics[width=0.2\textwidth]{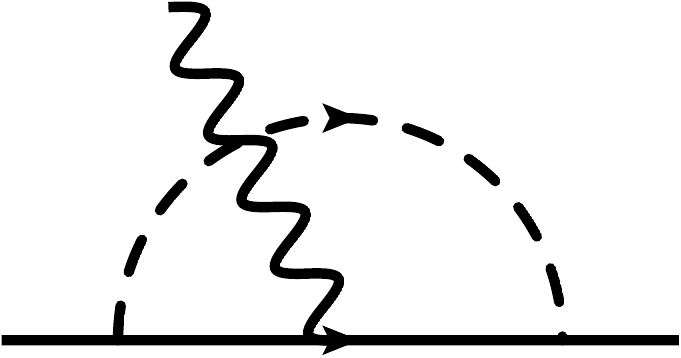}}
\subfigure[]{\label{Fdiagschptg}
\includegraphics[width=0.2\textwidth]{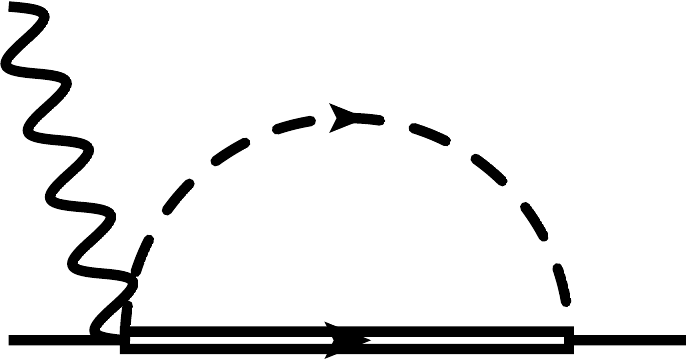}}
\subfigure[]{\label{Fdiagschpth}
\includegraphics[width=0.2\textwidth]{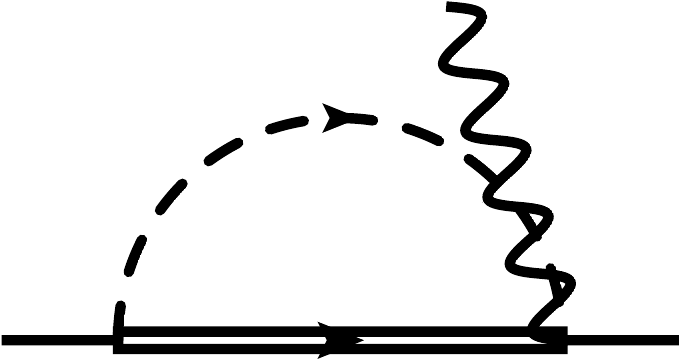}}
\subfigure[]{\label{Fdiagschpti}
\includegraphics[width=0.2\textwidth]{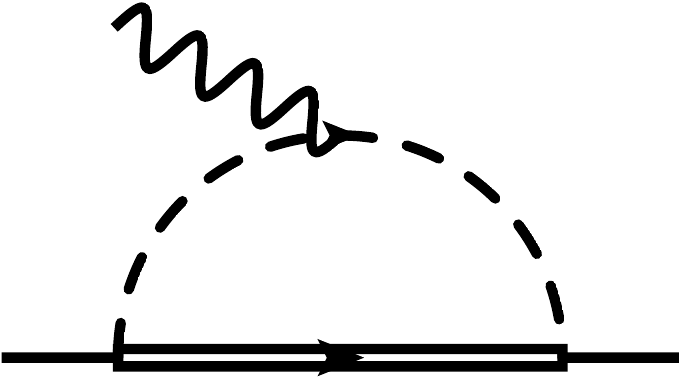}}
\subfigure[]{\label{Fdiagschptj}
\includegraphics[width=0.2\textwidth]{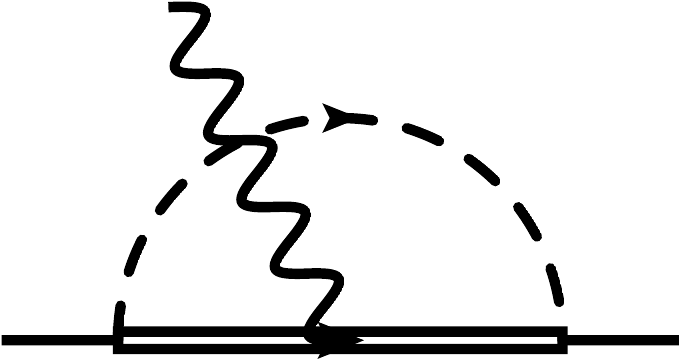}}
\caption{Loop diagrams that contribute to the baryon form factors up to $\mathcal{O}(p^3)$. The wavy, dashed, and single (double) continuous lines correspond to photon, pseudoscalar meson, and octet (decuplet) baryon fields, respectively. All the vertices are of the leading-order Lagrangians.}
\label{Fdiagschpt}
\end{figure}
\end{center}

The diagrams contributing up to chiral order $\mathcal{O}(p^3)$ are depicted in \cref{Fdiagschpttree,Fdiagschpt}. The expressions for the amplitudes obtained from these figures are discussed in the appendices. We show the wave-function renormalization (WFR) in App.~\ref{SappWFR}, where it is discussed that its inclusion is crucial in order to obtain the correct behavior at $q^2=0$. In App.~\ref{Sappamp}, we give the explicit expressions for the amplitudes of each diagram, decomposed into the form factors. Finally, in App.~\ref{SappPCBT}, we discuss the PCBTs that have to be subtracted. 


\subsection{Vector-meson contributions}\label{SformSvec}
\begin{center}
\begin{figure}
\includegraphics[width=0.2\textwidth]{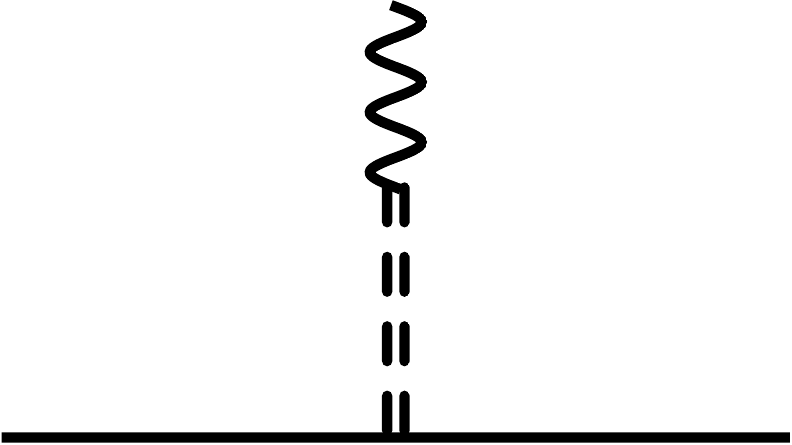}
\caption{Diagram with vector-meson contributions to the baryon form factors. The wavy, double dashed, and continuous lines correspond to photon, vector meson, and octet-baryon fields, respectively.}
\label{Fdiagsvec}
\end{figure}
\end{center}
In order to model the behaviour of the form factors at higher momentum transfers, the contributions of the vector mesons are also included, as has also been done in works such as Refs.~\cite{Kubis:2000aa,Kubis:2000zd}. The corresponding diagram to the order considered here is shown in \cref{Fdiagsvec}. The Lagrangian describing the couplings of the vector-meson fields $V_\mu$ with momentum $q$ to the octet baryons is given by~\cite{Drechsel:1998hk}
\begin{align}
\mathcal{L}_{VNN}=&\bar{B}\left(g_v\gamma^\mu+g_t\frac{\mathrm{i}\sigma^{\mu\nu}q_\nu}{2m_{B0}}\right)V_\mu B,
\end{align}
where $g_i\in \{g_v,g_t\}$ are the coupling constants which for the different baryons are related in $SU(3)$ as follows:
\begin{equation}
g_i = g_i^F~\text{Tr}(\bar B [V_8, B]) 
 +  g_i^D~\text{Tr}(\bar B \{ V_8, B \} )+ g_i^S~V_1 \text{Tr} (\bar B B ).\label{EqVecCoup}
\end{equation}
The explicit matrix representation of the octet and singlet vector fields $V_{8,1}^\mu$ is given e.g.\ in Refs.~\cite{Alarcon:2017asr,HillerBlin:2016jpb}.
%
%
In fact, in Ref.~\cite{Unal:2015hea} it has been shown with a Dirac constraint analysis that $g_v^D$ has to vanish.

We assume the case of ideal mixing, where the mixing angle $\varphi$ between the $\omega$ and the $\phi$ is such that $\sin\varphi=1/\sqrt{3}$.
%
%
The Lagrangian coupling a photon to the vector mesons $V_{\mu\nu}=\partial_\mu V_\nu-\partial_\nu V_\mu$ is given by~\cite{Borasoy:1995ds}
\begin{align}
\mathcal{L}_{V\gamma}=-\frac{1}{2\sqrt{2}}\frac{F_V}{m_V}\text{Tr}\left(V_{\mu\nu}f_+^{\mu\nu}\right),
\end{align}
where the mass and the decay constant of the vector mesons are given by $m_V$ and $F_V$, respectively. In \cref{tab:vecmesparams}, the values for the masses are shown, as well as the values for the decay constants resulting from the respective decay widths $\Gamma_{V\rightarrow e^+e^-}$~\cite{Agashe:2014kda}, via the correspondence
\begin{equation}
\Gamma_{V\rightarrow e^+e^-}=\frac{e^4F_V^2|\vec{p}|}{6\pi m_V^2}\approx\frac{4\pi\alpha^2F_V^2}{3m_V}.
\end{equation}
In the last step, the electron masses were neglected compared to the vector-meson mass, thus leading to $|\vec{p}|\approx m_V/2$. One can see that the numerical values for the $F_V$ are in good agreement with $SU(3)$ symmetry (the isospin couplings are also shown in the table), since they are of very similar size.
\begin{table}
\footnotesize
\begin{tabular}{l ccc}
\toprule
$V$&$m_V$~[MeV]&$F_V$~[MeV]&Is$_{V}$\\
\midrule
$\rho$&$775$&$156$&1\\
$\omega$&$783$&$138$&$\frac{1}{3}$\\
$\phi$&$1019$&$161$&$\frac{\sqrt{2}}{3}$\\
\bottomrule
\end{tabular}
\caption{Masses of the vector mesons, their decay constants, and values of the isospin coupling to the photon field.}
\label{tab:vecmesparams}
\end{table}
%
%

From Eq.~\eqref{EqVecCoup}, one can extract the $g_i$ for all the couplings of the baryons to the vector mesons, listed in \cref{tab:givec}, 
and relate them to the empirical couplings from nucleon-nucleon scattering data. In the present work, those from Ref.~\cite{Chiang:2001as} are used. Note, however, that the values are model dependent and have large uncertainties, compare, e.g.\ with Refs.~\cite{Machleidt:1987hj,Machleidt:2000ge}. The values we use for our calculations read
\begin{align}
g_{v,\rho^0 pp}=2.4,\quad g_{v,\omega pp}=16,\quad g_{t,\rho^0 pp}=14.6,\quad g_{t,\omega pp}=0.
\end{align}
Furthermore, assuming that the electromagnetic couplings of the baryons are saturated by vector-meson dominance, and considering the ratios of their electric charges and magnetic moments with the assumption of $SU(6)$ symmetry, one gets
\begin{align}
\frac{g_{v}^F}{ g_{v}^F + g_{v}^D } = 1,\quad \frac{g_{v}^F+g_{t}^F}{ g_{v}^F + g_{v}^D + g_{t}^F + g_{t}^D } = \frac{2}{5},
\end{align}
respectively~\cite{Dover:1985ba}. Therefore, one can extract the following information, with the help of which one can obtain all the other couplings between the vector mesons and the octet baryons:
\begin{align}
 g_{v}^F = 3.4, \quad g_{v}^D = 0,\quad g_{v}^S = 16.7,\quad
 g_{t}^F = 6.2, \quad g_{t}^D = 14.5,\quad g_{t}^S = -1.2.
\end{align}
Note that this is indeed compatible with the outcome of Ref.~\cite{Unal:2015hea}, where from theoretical principles $g_v^D=0$ was obtained.
\begin{table}
\footnotesize
    \begin{tabular}{l ccc}
\toprule
&$\rho$&$\omega$&$\phi$\\
\midrule
$p$&$\frac{g_i^F+g_i^D}{\sqrt{2}}$&$\frac{3g_i^F-g_i^D+2\sqrt{3}g_i^S}{3\sqrt{2}}$&$-\frac{g_i^D-3g_i^F+\sqrt{3}g_i^S}{3}$\\
$n$&$-\frac{g_i^F+g_i^D}{\sqrt{2}}$&$\frac{3g_i^F-g_i^D+2\sqrt{3}g_i^S}{3\sqrt{2}}$&$-\frac{g_i^D-3g_i^F+\sqrt{3}g_i^S}{3}$\\
$\Sigma^+$&$\sqrt{2}g_i^F$&$\frac{\sqrt{2}g_i^D+\sqrt{6}g_i^S}{3}$&$\frac{2g_i^D-\sqrt{3}g_i^S}{3}$\\
$\Sigma^0$&$0$&$\frac{\sqrt{2}g_i^D+\sqrt{6}g_i^S}{3}$&$\frac{2g_i^D-\sqrt{3}g_i^S}{3}$\\
$\Sigma^-$&$-\sqrt{2}g_i^F$&$\frac{\sqrt{2}g_i^D+\sqrt{6}g_i^S}{3}$&$\frac{2g_i^D-\sqrt{3}g_i^S}{3}$\\
$\Lambda$&$0$&$-\frac{\sqrt{2}g_i^D-\sqrt{6}g_i^S}{3}$&$-\frac{2g_i^D+\sqrt{3}g_i^S}{3}$\\
$\Xi^0$&$\frac{g_i^F-g_i^D}{\sqrt{2}}$&$-\frac{3g_i^F+g_i^D-2\sqrt{3}g_i^S}{3\sqrt{2}}$&$-\frac{g_i^D+3g_i^F+\sqrt{3}g_i^S}{3}$\\
$\Xi^-$&$\frac{g_i^D-g_i^F}{\sqrt{2}}$&$-\frac{3g_i^F+g_i^D-2\sqrt{3}g_i^S}{3\sqrt{2}}$&$-\frac{g_i^D+3g_i^F+\sqrt{3}g_i^S}{3}$\\
\bottomrule
    \end{tabular}
  \caption{Values of the isospin constants $g_i$ for the couplings of vector mesons to the octet baryons.}
  \label{tab:givec}
\end{table}


\section{Results}\label{Sres}
\subsection{Magnetic moments}

It is convenient to first calculate the observable $G_M(0)$, which is nothing other than the magnetic moment in units of the nuclear magneton, since the only unknown parameters that it depends on are $b_D$ and $b_F$. Moreover, for the neutron there is no dependence on $b_F$, and its experimental value is well determined to be $G_M^n(0)=-1.913$~\cite{Agashe:2014kda}. Therefore, one extracts the value $b_D=3.82$ when using the numerical values for the masses and other constants as summarized in \cref{Tab:chiparams}, following Ref.~\cite{Ledwig:2014rfa}. Using this result, one then extracts $b_F=0.97$ from the also well-determined experimental value for the proton, $G_M^p(0)=2.79$.
\begin{table}
\footnotesize
    \begin{tabular}{l cccccccc}
\toprule
                   $M_{B0}$  & $M_\Delta$ & $m_\pi$ & $m_K$ & $m_\eta$ & $F_0$  & $D$    & $F$    & $\mathcal{C}$\\
\midrule
      $880$  & $1152$   & $140$  & $496$ & $547$   & $87$  & $0.623$ & $0.441$ & $-D$\\
\bottomrule
    \end{tabular}
  \caption{Numerical values for the hadron masses and low-energy constants used in the calculations~\cite{Ledwig:2014rfa}. All the dimensionful values are given 
    in units of MeV.}
  \label{Tab:chiparams}
\end{table}

With these two constants fixed, one can now give predictions for $G_M(0)$ and the magnetic radius squared $\left<r^2_M\right>$ of all the baryon-octet members. The values are summarized in \cref{Tab:res_GM0,Tab:res_rM2}, and compared to the experimental~\cite{Agashe:2014kda} and lattice results~\cite{Shanahan:2014uka} where available.
\begin{table}
\footnotesize
    \begin{tabular}{l dddd}
\toprule
                    $G_M(0)$ & \multicolumn{1}{c}{$p$}  & \multicolumn{1}{c}{$n$} & \multicolumn{1}{c}{$\Sigma^+$} & \multicolumn{1}{c}{$\Sigma^-$} \\
\midrule
      This work&2.79  & -1.913   & 2.1(4)  & -1.1(1)\\
      This work (no $\Delta$)&2.79  & -1.913   & 2.5(2)  & -1.4(1)\\
      Experiment~\cite{Agashe:2014kda}&2.79  & -1.913 & 2.458(10)   & -1.160(25)\\
      Lattice~\cite{Shanahan:2014uka}&2.3(3)  & -1.45(17) & 2.12(18)   & -0.85(10) \\ 
      $\mathcal{O}(p^3)$~\cite{Kubis:2000aa} (no $\Delta$)&2.61  & -1.69 & 2.53   & -1.160\\
      $\mathcal{O}(p^4)$~\cite{Kubis:2000aa} (no $\Delta$)&2.79  & -1.913 & 2.458   & -1.00\\
      $\mathcal{O}(p^3)$~\cite{Geng:2009hh} (no VM)&2.61  & -2.23 & 2.37   & -1.17\\
      $\mathcal{O}(p^3)$ HBChPT~\cite{Bernard:1998gv}&2.79  & -1.913 & 2.8(4)   & -0.9(1)\\
      QM~\cite{Liu:2013fda}&2.735(121)  & -1.956(103) & 2.537(201)   & -0.861(40)\\
      NJL~\cite{Carrillo-Serrano:2016igi}&2.78  & -1.81 & 2.62   & -1.62 \\
\bottomrule\\
\toprule
                    $G_M(0)$ &\multicolumn{1}{c}{$\Sigma^0$} & \multicolumn{1}{c}{$\Lambda$}  & \multicolumn{1}{c}{$\Xi^0$}    & \multicolumn{1}{c}{$\Xi^-$}  \\ 
\midrule
      This work& 0.5(2)   & -0.5(2)  & -1.0(4) & -0.7(1)\\
      This work (no $\Delta$)& 0.6(2)   & -0.6(2)  & -1.1(3) & -0.98(2)\\
      Experiment~\cite{Agashe:2014kda}& \none & -0.613(4) & -1.250(14) & -0.6507(25)\\
      Lattice~\cite{Shanahan:2014uka}& \none & \none & -1.07(7) & -0.57(5) \\
      $\mathcal{O}(p^3)$ (no $\Delta$)~\cite{Kubis:2000aa}&0.76  & -0.76 & -1.51   & -0.93  \\
      $\mathcal{O}(p^4)$ (no $\Delta$)~\cite{Kubis:2000aa}&0.649  & -0.613 & -1.250   & -0.651  \\
      $\mathcal{O}(p^3)$ (no VM)~\cite{Geng:2009hh}&0.60  & -0.60 & -1.22   & -0.92\\
      $\mathcal{O}(p^3)$ HBChPT~\cite{Bernard:1998gv}&1.0(2)  & -1.0(2) & -1.9(4)   & -0.9(1)\\
      QM~\cite{Liu:2013fda}&0.838(91)  & -0.867(74) & -1.690(142)   & -0.840(87)   \\
      NJL~\cite{Carrillo-Serrano:2016igi}& \none  & \none & -1.14   & -0.67 \\
\bottomrule
    \end{tabular}
  \caption{Numerical values for $G_M(0)$, compared with those extracted experimentally~\cite{Agashe:2014kda} and on the lattice~\cite{Shanahan:2014uka}. We compare our full model with a refitted version without the inclusion of the decuplet ($\Delta$) states. We also show the $\mathcal{O}(p^3)$ and $\mathcal{O}(p^4)$ ChPT calculations of Ref.~\cite{Kubis:2000aa}, which does not include the decuplet (no $\Delta$) intermediate states, the $\mathcal{O}(p^3)$ ChPT calculation which does not include the vector mesons (no VM)~\cite{Geng:2009hh}, a refit of the HBChPT results of Ref.~\cite{Bernard:1998gv} for $SU(3)$, a quark-model (QM) approach~\cite{Liu:2013fda}, and a calculation within the Nambu--Jona--Lasinio (NJL) model~\cite{Carrillo-Serrano:2016igi}.}
  \label{Tab:res_GM0}
\end{table}

For $G_M(0)$, since the experimental errors are negligible, the main uncertainties arise from the choice of values for the parameters used in our calculation: e.g.\ if we were to use the physical average for the masses, decay constants and other coupling constants, the final values for $b_D$ and $b_F$ would change by $5\%$ and $15\%$, respectively, when maintaining fixed the values for $G_M^p(0)$ and $G_M^n(0)$. This leads to errors for the final results for $G_M(0)$ as shown in \cref{Tab:res_GM0}. We find that in our approach the results are in very good agreement with those found experimentally, and compatible with those obtained on the lattice. This shows that despite $SU(3)$-symmetry breaking, the framework of $SU(3)$ ChPT gives reliable results. The prediction for $G_M^\Lambda(0)$ turns out to be of the same magnitude as $G_M^{\Sigma^0}$, but with opposite sign. The same behaviour has been seen in other approaches~\cite{Kubis:2000aa,Geng:2009hh,Liu:2013fda}.

Our approach builds upon that of Ref.~\cite{Geng:2009hh}. There, the renormalization scheme used was also EOMS, and the spin-3/2 states were included as explicit intermediate states too. The vector-meson contributions were not included, but they do not enter the result of $G_M(0)$ since the contribution of vector mesons at $q^2=0$ vanishes. Thus, in that work they gave a complete analysis of the numerical values for this particular observable. The difference in our results lies in the fact that they did a global fit to all the experimentally extracted magnetic moments available, while we fixed our results to those of the best determined ones: those of the nucleons.

We show the results presented in Ref.~\cite{Kubis:2000aa}, which were also obtained within the framework of covariant ChPT. The main differences between their work and ours are the renormalization scheme used (they use IR instead of EOMS), the order calculated (in that work the calculations were performed up to $\mathcal{O}(p^4)$), and the fact that they did not consider the explicit inclusion of the spin-3/2 intermediate states. The values we obtained for $b_D$ and $b_F$ are compatible with those obtained in Ref.~\cite{Kubis:2000aa} at $\mathcal{O}(p^3)$, keeping in mind that they have to be slightly different due to the different renormalization schemes and intermediate states considered.

In fact, in \cref{Tab:res_GM0} we also show the results that were obtained in Ref.~\cite{Kubis:2000aa} at order $\mathcal{O}(p^3)$, and our refitted results at the same order without the inclusion of the decuplet. Thus, the only difference is the renormalization scheme used and the results for the observables should be equivalent. This is indeed the case, and it is important to stress that the data for the nucleon magnetic moments used in Ref.~\cite{Kubis:2000aa} are outdated. Therefore the fit results can of course not be compared exactly.

It is interesting to see that for the observable $G_M(0)$ it suffices to stay at $\mathcal{O}(p^3)$ if one includes the decuplet intermediate states explicitly, as was done in the present work. The results turn out to be as good the $\mathcal{O}(p^4)$ calculation in Ref.~\cite{Kubis:2000aa}, despite being of a lower order. In fact, while for the neutron the contribution of the intermediate decuplet states is negligible, in the other cases these intermediate states make up $20$ to $90\%$ of the values, the most striking case being that of the $\Xi^-$. This is reflected in the result for its magnetic moment, thus showing the importance of including the spin-3/2 states in order to obtain a result compatible with the experiment. We note that our agreement with the data seems to be overall slightly better than that found in some other approaches, such as relativistic quark models~\cite{Liu:2013fda} and Nambu--Jona--Lasinio calculations~\cite{Carrillo-Serrano:2016igi}. Furthermore, the LECs appear to be more stable when including the decuplet: $b_D$ almost does not change, $b_F$ gets reduced to approximately half. In contrast, in Ref.~\cite{Kubis:2000aa}, when going to $\mathcal{O}(p^4)$, $b_D$ shifts from $3.65$ to $5.18$, and $b_F$ by more than a factor 3, from $1.73$ to $0.56$.

In order to get an idea about the dependence of the results on the renormalization used, it is useful to calculate the observables in other schemes as well. As discussed in the introduction, the IR scheme has the downside of failing to conserve the analytical structure of the loop diagrams. Furthermore, the EOMS scheme is technically much simpler to perform. Therefore, we refrain from comparing our EOMS results to a technically expensive calculation in the IR scheme.

Instead, we reanalyse the $SU(2)$ HBChPT calculations performed in Ref.~\cite{Bernard:1998gv}, by extending them to $SU(3)$. In order to do so, we use the expressions listed in that work, taking into account that the LECs' renormalization and the WFR have to be adjusted to the $SU(3)$ case studied in the present work. The spin-3/2 degrees of freedom had already been included in the original calculation. The result of the refit for all the octet baryon magnetic moments is shown in \cref{Tab:res_GM0}. One can see that, when fixing the LECs in order to reproduce the experimental values for the proton and the neutron, the hyperon magnetic moments seem to be slightly overestimated, with the exception of the $\Sigma^-$. The results qualitatively point into the correct direction, though. Here too one can find that the magnetic moments of the $\Sigma^0$ and the $\Lambda$ are the same up to their sign.

This comparative study confirms the need to perform covariant ChPT, and gives an idea about the dependence on the renormalization scheme used. As has already been suggested in Ref.~\cite{Gasser:2015dwa}, the fully covariant approach is expected to give a better convergence of the form factor results up to higher values of momentum transfer squared. Thus, it might be of interest to reanalyse the polarizabilities studied in Ref.~\cite{Gasser:2015dwa} with the calculations presented here.

\subsection{Magnetic radii}
The observable $\left<r_M^2\right>$ is slightly more involved: it depends on the effects of the intermediate vector mesons, the inclusion of which is model dependent. We chose the parameterization of Ref.~\cite{Chiang:2001as} for the vector-meson couplings, but when choosing the Bonn-potential values the results vary by $15$ to $40\%$ percent. Thus, for this observable the vector mesons are the main source of the uncertainty in the final results shown in \cref{Tab:res_rM2}. Taking this into account, our calculation is in good agreement with the experimental data for the nucleons.
\begin{table}
\footnotesize
    \begin{tabular}{l dddd}
\toprule
                    $\left<r_M^2\right>$~[fm$^2$] &  \multicolumn{1}{c}{$p$}  &  \multicolumn{1}{c}{$n$} &  \multicolumn{1}{c}{$\Sigma^+$} &  \multicolumn{1}{c}{$\Sigma^-$} \\
\midrule
      This work&0.9(2)  & 0.8(2)   & 1.2(2)  & 1.2(2)\\
      This work (no $\Delta$)&0.7(2)  & 0.7(3)   & 0.8(1)  & 0.7(2)\\
      Experiment~\cite{Agashe:2014kda}&0.777(16)  & 0.862(9) & \none   & \none\\
      Lattice~\cite{Shanahan:2014uka}&0.71(8)  & 0.86(9) & 0.66(5)   & 1.05(9)  \\
      $\mathcal{O}(p^4)$~\cite{Kubis:2000aa} (no $\Delta$)&0.699  & 0.790  & 0.80(5) & 1.20(13) \\
      $\mathcal{O}(p^3)$ HBChPT~\cite{Bernard:1998gv}&0.9(2)  & 1.0(2) & 0.8(2)   & 1.1(2)\\
      QM~\cite{Liu:2013fda}&0.909(84)  & 0.922(79) & 0.885(94)   & 0.951(83)  \\
      NJL~\cite{Carrillo-Serrano:2016igi}&0.87  & 0.91 & 0.88   & 0.96  \\
\bottomrule\\
\toprule
                    $\left<r_M^2\right>$~[fm$^2$] & \multicolumn{1}{c}{$\Sigma^0$} &  \multicolumn{1}{c}{$\Lambda$}  &  \multicolumn{1}{c}{$\Xi^0$}    &  \multicolumn{1}{c}{$\Xi^-$}\\
\midrule
      This work& 1.1(2)   & 0.6(2)  & 0.7(3) & 0.8(1)\\
      This work (no $\Delta$)& 0.8(2)   & 0.3(3)  & 0.5(1) & 0.2(1)\\
      Experiment~\cite{Agashe:2014kda}& \none & \none & \none & \none\\
      Lattice~\cite{Shanahan:2014uka}& \none & \none & 0.53(5) & 0.44(5) \\
      $\mathcal{O}(p^4)$~\cite{Kubis:2000aa} (no $\Delta$)&0.20(10)  & 0.48(9)   & 0.61(12)   & 0.50(16)\\
      $\mathcal{O}(p^3)$ HBChPT~\cite{Bernard:1998gv}&0.6(2)  & 0.3(2) & 0.4(3)   & 0.2(1)\\
      QM~\cite{Liu:2013fda}&0.851(102)  & 0.852(103) & 0.871(99)   & 0.840(109) \\
      NJL~\cite{Carrillo-Serrano:2016igi}& \none & \none & 0.66   & 0.51 \\
\bottomrule
    \end{tabular}
  \caption{Numerical values for $\left<r_M^2\right>$, compared with those extracted experimentally~\cite{Agashe:2014kda} and on the lattice~\cite{Shanahan:2014uka}. We compare our full model with a refitted version without the inclusion of the decuplet ($\Delta$) states. We also show the $\mathcal{O}(p^3)$ and $\mathcal{O}(p^4)$ ChPT calculations of Ref.~\cite{Kubis:2000aa}, which does not include the decuplet intermediate states (no $\Delta$), a refit of the HBChPT results of Ref.~\cite{Bernard:1998gv} for $SU(3)$, a quark-model (QM) approach~\cite{Liu:2013fda}, and a calculation within the Nambu--Jona--Lasinio (NJL) model~\cite{Carrillo-Serrano:2016igi}.}
  \label{Tab:res_rM2}
\end{table}

Concerning the hyperons, there are no experimental data available on their magnetic radii. Thus we compare our results to those extracted in other theoretical frameworks. We have good agreement with other predictions, although for some cases, such as the $\Sigma^+$, the $\Sigma^0$ and the the $\Xi^-$, the tendency is that our prediction is of a slightly larger magnetic radius. In those cases, there is a big difference between our result and that in Ref.~\cite{Kubis:2000aa}. As a result, our prediction for the magnetic radius of the $\Sigma^+$ is the same as that for $\Sigma^-$. Nevertheless, we would like to stress here that it has already been shown in Ref.~\cite{Kubis:2000aa} that it is crucial to perform a full $\mathcal{O}(p^4)$ calculation in order to obtain a good description of the $q^2$ dependence of the form factors $G_E(q^2)$ and $G_M(q^2)$. Since the charge and magnetic radii are obtained from the slope of the form factors at $q^2=0$, for these observables the effects of the next chiral order are felt more strongly than for the magnetic moment. It is also important to point out that the $\mathcal{O}(p^4)$ calculation performed in Ref.~\cite{Kubis:2000aa} was in perfect agreement with the experimental data existing at the time. It is intriguing, though, that the quark-model predictions~\cite{Liu:2013fda} are compatible with our result. In particular, the effect of the decuplet intermediate states is again striking in the final numerical result of the $\Xi^-$.

For this observable, too, we perform a comparative study in HBChPT, following Ref.~\cite{Bernard:1998gv}. Apart from extending their results to $SU(3)$, as explained in the previous section for the observable $G_M(0)$, we additionally include the vector-meson contributions, in order to obtain a direct comparison to our covariant model. Since the vector mesons enter at tree level only, their effect is not renormalization-scheme dependent, and we include them in exactly the same phenomenological way as in our fully covariant model. Again, the qualitative behaviour of the magnetic radii is compatible in the covariant and non-relativistic schemes, but the numerical values vary. This is especially true for the $\Xi^-$ of which the central value changes dramatically despite having small error bars.

\subsection{Charge radii}
Lastly, we show the results for the average charge radii squared, $\left<r_E^2\right>$, shown in \cref{Tab:res_rE2}. Again, the values are compared to the data and calculations in other works. We fixed $b_D$ and $b_F$ to the values extracted above, and determined $d_{101}$ and $d_{102}$ by comparison to the two experimental values for the proton and for the $\Sigma^-$. We obtained $d_{101}=0.61$ and $d_{102}=-0.70$, while a fit without the decuplet intermediate states would have resulted in $d_{101}=0.54$ and $d_{102}=-1.05$. The constant $d_{102}$ seems to be sensitive to the inclusion of the decuplet. However, the effect on $d_{101}$ is very soft, in contrast to the result of going to $\mathcal{O}(p^4)$, where $d_{101}$ changes by a factor $1/3$.
\begin{table}
\footnotesize
    \begin{tabular}{l dddd}
\toprule
                    $\left<r_E^2\right>$~[fm$^2$] & \multicolumn{1}{c}{$p$}  & \multicolumn{1}{c}{$n$} & \multicolumn{1}{c}{$\Sigma^+$} & \multicolumn{1}{c}{$\Sigma^-$} \\
\midrule
      This work&0.878  & 0.03(7)   & 0.99(3)  & 0.780\\
      This work (no $\Delta$)&0.878  & 0.04(7)   & 0.95(3)  & 0.780\\
      Experiment~\cite{Agashe:2014kda}&0.878(5)  & -0.1161(22) & \none   & 0.780(10)\\
      Lattice~\cite{Shanahan:2014cga}&0.76(10)  & \none &0.61(8) & 0.45(3)\\
      $\mathcal{O}(p^3)$~\cite{Kubis:2000aa} (no $\Delta$)&0.717  & -0.113 &0.63 & 0.72 \\
      $\mathcal{O}(p^4)$~\cite{Kubis:2000aa} (no $\Delta$)&0.717  & -0.113 &0.60(2) & 0.67(3) \\
      $\mathcal{O}(p^3)$ HBChPT~\cite{Bernard:1998gv}&0.878  & -0.04(7) & 0.93(3)   & 0.780\\
      QM~\cite{Liu:2013fda}&0.767(113)  & -0.014(1) & 0.781(108)   & 0.781(108) \\
      NJL~\cite{Carrillo-Serrano:2016igi}&0.87  & -0.37 & 0.96   & 0.86  \\
\bottomrule\\
\toprule
                    $\left<r_E^2\right>$~[fm$^2$] &\multicolumn{1}{c}{$\Sigma^0$} & \multicolumn{1}{c}{$\Lambda$}  & \multicolumn{1}{c}{$\Xi^0$}    & \multicolumn{1}{c}{$\Xi^-$} \\
\midrule
      This work& 0.10(2)   & 0.18(1)  & 0.36(2) & 0.61(1)\\
      This work (no $\Delta$)& 0.09(1)   & 0.20(2)  & 0.38(2) & 0.605(7)\\
      Experiment~\cite{Agashe:2014kda}& \none & \none & \none & \none\\
      Lattice~\cite{Shanahan:2014cga}& \none & \none & 0.53(5)    & 0.37(2)\\
      $\mathcal{O}(p^3)$~\cite{Kubis:2000aa} (no $\Delta$)& -0.05 & 0.05 & 0.15    & 0.56\\
      $\mathcal{O}(p^4)$~\cite{Kubis:2000aa} (no $\Delta$)& -0.03(1) & 0.11(2) & 0.13(3)    & 0.49(5)\\
      $\mathcal{O}(p^3)$ HBChPT~\cite{Bernard:1998gv}&0.07(2)  & 0.21(1) & 0.42(2)   & 0.54(1)\\
      QM~\cite{Liu:2013fda}&0  & 0 & 0.014(8)   & 0.767(113)   \\
      NJL~\cite{Carrillo-Serrano:2016igi}& \none & \none & 0.49   & 0.76 \\
\bottomrule
    \end{tabular}
  \caption{Numerical values for $\left<r_E^2\right>$, compared with those extracted experimentally~\cite{Agashe:2014kda} and on the lattice~\cite{Shanahan:2014cga}. We compare our full model with a refitted version without the inclusion of the decuplet ($\Delta$) states. We also show the $\mathcal{O}(p^4)$ ChPT calculation of Ref.~\cite{Kubis:2000aa}, which does not include the decuplet intermediate states (no $\Delta$), a refit of the HBChPT results of Ref.~\cite{Bernard:1998gv} for $SU(3)$, a quark-model (QM) approach~\cite{Liu:2013fda}, and a calculation within the Nambu--Jona--Lasinio (NJL) model~\cite{Carrillo-Serrano:2016igi}.}
  \label{Tab:res_rE2}
\end{table}

With the values of the LECs obtained from the fit, we gave predictions for the charge radii of the other members of the baryon octet. The findings here are similar to those remarked in the case of the magnetic radii: the results are sensitive to higher-order contributions, but overall a behaviour compatible with other calculations can be seen. Again, the choice between different vector-meson parameterizations is the main source of uncertainty in the results. Note that the data available at the time of the calculations performed in Refs.~\cite{Kubis:2000aa,Liu:2013fda} are outdated, which naturally leads to a discrepancy in the fit results. In the case of the charge radii, the inclusion of the decuplet is negligible, and the vector mesons are the ones giving significant contributions.

The model in Ref.~\cite{Carrillo-Serrano:2016igi} describes the nucleon data rather well, but the experimental value for $\Sigma^-$ is not reproduced. In our work, we find that when fixing the couplings so as to also reproduce the results for this baryon, we obtain a charge radius which is slightly larger for the $\Sigma^+$ than for the  $\Sigma^-$. This is not surprising, since the main contribution from the kaon cloud to the $\Sigma^+$ comes from the transition to a virtual $p~\bar{K^0}$ state, while that to the $\Sigma^-$ is $n~K^-$. Thus, while the pion cloud contributes equally to both $\Sigma^+$ and $\Sigma^-$, the kaon cloud leads to a breaking of this symmetry. This is an interesting outcome of the extension of the ChPT calculations to $SU(3)$. Additionally, the relative sign between the contributions of the $\rho$ and the other two vector mesons, $\omega$ and $\phi$, is different depending on the charge of the $\Sigma$. Finally, even when considering the direct coupling of the photon to these baryons at $\mathcal{O}(p^2)$ and $\mathcal{O}(p^3)$, one finds different contributions to each of them, since the relative sign between the charge $c_b$ and $c_{b23}$ is different for each case, see \cref{tab:cb23}.

Unlike the magnetic observables, the charge radii do not change drastically when performing a HBChPT calculation instead of the fully covariant approach. Again, the hyperon of which the central value changes the most compared to its error bars is the $\Xi^-$, but not as strongly as in the case of the magnetic radius. Furthermore, it is important to point out that in the HBChPT calculation the central value for the neutron is indeed negative, as expected from the experiment and other calculations.


\section{Summary}\label{Ssum}

We presented a systematic and extensive calculation of the baryon electromagnetic form factors within the framework of covariant ChPT up to the chiral order $\mathcal{O}(p^3)$. Building on the work of Ref.~\cite{Geng:2009hh}, we explicitly included not only the decuplet, but also the vector-meson contributions in the covariant EOMS renormalization scheme. In addition to the magnetic moments, we analysed the charge and magnetic radii.

We first introduced the tools necessary for the calculation of the relevant amplitudes. With this framework we extracted the magnetic moments, charge and magnetic radii. Comparing the results with data, we determined all unknown low-energy constants. Finally, we provided predictions for the properties of those baryons for which these observables have not yet been determined experimentally.

Our results for the magnetic moments are in excellent agreement with the data. In fact, we even find as good an agreement as in calculations at higher chiral order $\mathcal{O}(p^4)$. This shows the importance of including the spin-3/2 degrees of freedom explicitly, and the reliability of $SU(3)$ ChPT in its covariant EOMS renormalization framework. Both in the case of the magnetic moment and of the magnetic radius, we find that the effect of the decuplet intermediate states is crucial for the result for the $\Xi^-$ hyperon.

However, we confirmed that for observables sensitive to the $q^2$ behaviour, such as the charge and magnetic radii, an extension of these calculations to the chiral order $\mathcal{O}(p^4)$ is crucial. So far, such calculations in $SU(3)$ have been performed only in the IR renormalization scheme and without the explicit inclusion of the decuplet intermediate states. However, even at $\mathcal{O}(p^3)$ we obtained results in good agreement with data. In order to study the effect on the results of the renormalization scheme used, we compared our calculations to those obtained in HBChPT. Since this scheme is non-relativistic, it is the one which is expected to lead to the most different numerical outcome. Thus, it gives a quantitative idea about the uncertainties due to different renormalization schemes.

Finally, we would like to stress that the proton charge radius extracted with the help of this framework by a fit to the $G_E(q^2)$ data at low $q^2$ might be more reliable than the polynomial fits that are usually performed. The latter cannot take into account the effects of poles in the amplitude, such as those at the opening of the two-pion threshold. Such effects are correctly described within ChPT.

\begin{acknowledgments}
This work was supported by the 
Spanish Ministerio de Econom\'{\i}a y Competitividad (MINECO) and the European fund for regional development (EFRD) under contracts No.~FIS2014-51948-C2-2-P and No.~SEV-2014-0398. It has also been supported by Generalitat Valenciana under contract PROMETEOII/2014/0068 and the Deutsche Forschungsgemeinschaft DFG. A.N.H.B. thanks Jose Manuel Alarc\'on, Tim Ledwig, Vladimir Pascalutsa, Stefan Scherer, Zhi-Feng Sun, Manuel J.~Vicente Vacas, and Christian Weiss for valuable discussions.
\end{acknowledgments}

\begin{appendices}


\section{Wave-function renormalization}\label{SappWFR}
\begin{figure}[htbp]
\begin{center}
\subfigure[]{
\label{fSelfEna}
\includegraphics[width=0.2\textwidth]{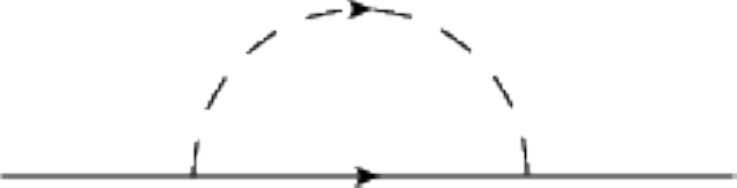}}
\subfigure[]{
\label{fSelfEnb}
\includegraphics[width=0.2\textwidth]{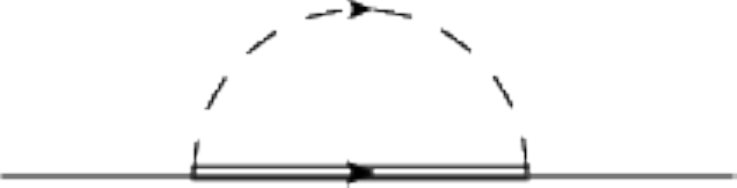}}
\end{center}
\caption{Diagrams contributing to the proton self-energy with nucleon (a) and $\Delta$ (b) loop contributions.}
\label{fSelfEn}
\end{figure}

At the order considered, the WFR amounts to including a factor $\sqrt{Z}$ in the amplitude, for each of the baryon legs. The WFR of the photon leg would lead to higher-order corrections. Therefore, in total one needs
\begin{equation}\label{wfr}
\sqrt{Z}^2 = \frac{1}{1-\Sigma'}\Big|_{\slashed p = m_{B0}},
\end{equation}
where $\Sigma$ is the baryon self-energy that arises from the diagrams depicted in \cref{fSelfEn}. This factor $Z$ is of $\mathcal{O}(p^2)$, and therefore gives a $\mathcal{O}(p^3)$ correction when included in the tree-level amplitude of $\mathcal{O}(p)$. Therefore, we include it only there, since in the higher-order diagrams one would obtain corrections of at least $\mathcal{O}(p^4)$.

In fact, this WFR is strictly necessary if one wants to obtain the natural charge of the baryons  at $q^2=0$:
\begin{equation}
G_E(0)= c_b.
\end{equation}
The reason for this is that the loop diagrams at $\mathcal{O}(p^3)$ do not all show the behaviour of $G_E(q^2=0)=0$, even after subtracting the PCBTs. But the expressions obtained for the WFR exactly cancel these spurious terms when multiplied to the leading-order tree-level diagram:
\begin{equation}
c_b\text{WFR}=c_b+G_E(0)^\text{loops}=c_b+F_{1}(0)^\text{loops}-F_{1}(0)^\text{PCBT}.\label{EqWFRcancel}
\end{equation} 
For the neutral baryons this of course means that there is no WFR contribution. But in their case the requirement $G_E(0)=0$ is fulfilled even without WFR. The WFR expression for the charged baryons is trivially obtained from \cref{EqWFRcancel,EqF1amps,EqPCBT1amps}, and therefore, we do not show it here.

\section{Explicit expressions for the form factors}\label{Sappamp}
In this section, the amplitudes of the diagrams in \cref{Fdiagschpttree,Fdiagschpt} are given, separated into the two form factors. The charges of the meson, of the octet baryon and of the decuplet baryon are denoted as $c_m$, $c_{b}$ and $c_T$, respectively. For the direct coupling of the photon to the baryon at $\mathcal{O}(p^2)$ and $\mathcal{O}(p^3)$, apart from the baryon charge a further definition $c_{b23}$ is needed, the values of which are summarized for each baryon in \cref{tab:cb23}. We denote the isospin constant of the coupling of two mesons to a baryon at one point as Is$_{mm}$. Its values for the different baryons are summarized in \cref{tab:IsmmPi,tab:IsmmKaon}. In some of the loop diagrams, each channel has a particular isospin combination of the LECs $D$ and $F$. Therefore, we call the combination thereof $c_{DF}$. The values of this combination for the different channels are summarized in \cref{tab:cdf,tab:cdf_eta}. Furthermore, Is$_m$ is the coupling constant of the vertex of the decuplet-to-octet baryon transition via a coupling to a meson. Its values are summarized in \cref{tab:Ism,tab:Ism_eta}. The couplings Is$_V$ between a photon and a vector meson are summarized in \cref{tab:vecmesparams}.

For the expressions that arise from the dimensional regularization we use the definitions
\begin{align}
\nn&\lambda_1(\Delta)=\frac{\Gamma\left(1-\frac d2\right)}{(4\pi)^{d/2}\Delta^{1-\frac d2}}= -\frac{\Delta}{16\pi^2}\left[\frac{2}{\epsilon}-\log\left(\frac{\Delta}{M_{\text{Sc}}}\right)+\log(4\pi)-\gamma_E+1+\mathcal{O}(\epsilon)\right],\\[1em]
\nn&\lambda_2(\Delta)=\frac{\Gamma\left(2-\frac d2\right)}{(4\pi)^{d/2}\Delta^{2-\frac d2}}= \frac{1}{16\pi^2}\left[\frac{2}{\epsilon}-\log\left(\frac{\Delta}{M_{\text{Sc}}}\right)+\log(4\pi)-\gamma_E+\mathcal{O}(\epsilon)\right],\\[1em]
\nn&\lambda_3(\Delta)=\frac{\Gamma\left(3-\frac d2\right)}{(4\pi)^{d/2}\Delta^{3-\frac d2}}
= \frac{1}{16\pi^2\Delta}+\mathcal{O}(\epsilon),\\[1em]
\nn&\rho_1(\Delta)=-\frac{\Delta}{8\pi^2},\\[1em]
&\rho_2(\Delta)= \frac{1}{8\pi^2},
\end{align}
where $\epsilon=4-d$ and $M_\text{Sc}$ is the scale parameter, which in this work is set to the octet-baryon mass $m_{B0}$. Furthermore, $\gamma_E=-\Gamma'(1)$ is the Euler-Mascheroni constant. In the renormalization prescription $\widetilde{\text{MS}}$ used here, terms proportional to
\begin{align}
L=\frac{2}{\epsilon} + \log(4\pi) - \gamma_E + 1
\end{align}
are subtracted. Special care has to be taken for amplitude terms which are proportional to the dimension $d = 4-\epsilon$. They arise, e.g.\ from expressions as $g^{\mu\nu}\gamma_\mu\gamma_\nu=d=4-\epsilon$ in the numerator. The $\epsilon$ piece of these expressions cancels the divergence in $2/\epsilon$, therefore leading to the appearance of the additional finite terms $\rho_1(\Delta)$ and $\rho_2(\Delta)$ which are not absorbed into the renormalization. Were one to set $d=4$ from the very beginning, they would have erroneously disappeared.

Furthermore, we define the following arguments of the loop integrals:
\begin{align}
\nn\Delta_{\ref{Fdiagschpta}}=&M^2,\\[1em]
\nn\Delta_{\ref{Fdiagschptb}}=&M^2-q^2f_a(1-f_a),\\[1em]
\nn\Delta_{\ref{Fdiagschptc}}=&\Delta_{\ref{Fdiagschptd}}=M^2(1-f_a)+f_a^2m_{B0}^2,\\[1em]
\nn\Delta_{\ref{Fdiagschpte}}=&M^2(1-f_b)+f_b^2m_{B0}^2-q^2f_a(1-f_a-f_b),\\[1em]
\nn\Delta_{\ref{Fdiagschptf}}=&M^2(1-f_a-f_b)+(f_a+f_b)^2m_{B0}^2-q^2f_af_b,\\[1em]
\nn\Delta_{\ref{Fdiagschptg}}=&\Delta_{\ref{Fdiagschpth}}=M^2(1-f_a)-m_{B0}^2f_a(1-f_a)+M_\Delta^2f_a,\\[1em]
\nn\Delta_{\ref{Fdiagschpti}}=&M^2(1-f_b)-m_{B0}^2f_b(1-f_b)+M_\Delta^2f_b-q^2f_a(1-f_a-f_b),\\[1em]
\Delta_{\ref{Fdiagschptj}}=&M^2(1-f_a-f_b)+(f_a+f_b)(M_\Delta^2-m_{B0}^2)+(f_a+f_b)^2m_{B0}^2-q^2f_af_b.
\end{align}

With the above considerations, the evaluation of the diagrams in \cref{Fdiagschpttree,Fdiagschpt} leads to the following expressions for the form factors $F_1$ and $F_2$:
\begin{align}
\nn F_{1,\ref{Fdiagschpttree}}&=c_b+q^2\left(c_bd_{101}+c_{b23}d_{102}\right),\\[1em]
\nn F_{1,\ref{Fdiagschpta}}&=-\frac{\text{Is}_{mm}}{F_0^2}\lambda_1\left(\Delta_{\ref{Fdiagschpta}}\right),\\[1em]
\nn F_{1,\ref{Fdiagschptb}}&=-2\frac{\text{Is}_{mm}}{F_0^2}\int_0^1{\mathrm{d}f_a}~\lambda_1\left(\Delta_{\ref{Fdiagschptb}}\right),\\[1em]
\nn F_{1,\ref{Fdiagschptc}}&=F_{1,\ref{Fdiagschptd}}=\frac{c_{DF}^2c_m}{4F_0^2}\int_0^1{\mathrm{d}f_a}~\Big[2\lambda_1\left(\Delta_{\ref{Fdiagschptc}}\right)
-m_{B0}^2f_a^2\lambda_2\left(\Delta_{\ref{Fdiagschptc}}\right)-\frac{1}{2}\rho_1\left(\Delta_{\ref{Fdiagschptc}}\right)\Big],\\[1em]
\nn F_{1,\ref{Fdiagschpte}}&=\frac{c_{DF}^2c_m}{4F_0^2}\int_0^1{\mathrm{d}f_a}\int_0^{1-f_a}{\mathrm{d}f_b}~\Big[-3\lambda_1\left(\Delta_{\ref{Fdiagschpte}}\right)\\[1em]
\nn&+\left(m_{B0}^2(4f_a(4f_b+5)+17f_b^2+8f_b-10)+q^2f_a(f_a+f_b-1)\right)\lambda_2\left(\Delta_{\ref{Fdiagschpte}}\right)\\[1em]
\nn&-2m_{B0}^2(2f_a+2f_b-1)\left[q^2\left(f_af_b^2+f_a(f_a+1)f_b+2(f_a-1)f_a\right)+f_b^3m_{B0}^2\right]\lambda_3\left(\Delta_{\ref{Fdiagschpte}}\right)\\[1em]
\nn&+\frac{1}{2}\rho_1\left(\Delta_{\ref{Fdiagschpte}}\right)
-m_{B0}^2(f_b+2)(2f_a+2f_b-1)\rho_2\left(\Delta_{\ref{Fdiagschpte}}\right)\Big],\\[1em]
\nn F_{1,\ref{Fdiagschptf}}&=\frac{c_{DF}^2c_b}{4F_0^2}\int_0^1{\mathrm{d}f_a}\int_0^{1-f_a}{\mathrm{d}f_b}~\Big[6\lambda_1\left(\Delta_{\ref{Fdiagschptf}}\right)\\[1em]
\nn&-\left(6m_{B0}^2(f_a^2+2f_a(f_b+1)+f_b^2-2f_b-1)+q^2(f_a(3-6f_b)+3f_b-1))\right)\lambda_2\left(\Delta_{\ref{Fdiagschptf}}\right)\\[1em]
\nn&+\Big(m_{B0}^4(f_a+f_b)^2(f_a^2+2f_a(f_b+2)+(f_b-4)f_b)\\[1em]
\nn&+m_{B0}^2\left(f_a^3(1-2f_b)-f_a^2f_b(4f_b+1)+f_af_b((7-2f_b)f_b+4)+f_b^3\right)q^2\\[1em]
\nn&+(f_a-1)f_a(f_b-1)f_bq^4\Big)\lambda_3\left(\Delta_{\ref{Fdiagschptf}}\right)\\[1em]
\nn&-\frac{5}{2}\rho_1\left(\Delta_{\ref{Fdiagschptf}}\right)
+\left(m_{B0}^2(f_a^2+2f_a(f_b+1)+f_b^2-2f_b-2)+\frac{q^2}{2}(-2f_af_b+f_a+f_b)\right)\rho_2\left(\Delta_{\ref{Fdiagschptf}}\right)\Big],\\[1em]
\nn F_{1,\ref{Fdiagschptg}}&=\frac{\mathcal{C}^2m_{B0}\text{Is}_{m}^2}{9M_\Delta^3F_0^2}\int_0^1{\mathrm{d}f_a}~
\left(f_am_{B0}-M_\Delta-m_{B0}\right)\\
\nn&\times\left\{6(M_\Delta(c_T + c_m) - 2 m_{B0} c_T)\lambda_1\left(\Delta_{\ref{Fdiagschptg}}\right)-(M_\Delta(c_T + c_m) - 8 m_{B0} c_T)\rho_1\left(\Delta_{\ref{Fdiagschptg}}\right)\right\},\\[1em]
\nn F_{1,\ref{Fdiagschpth}}&=-\frac{\mathcal{C}^2m_{B0}\text{Is}_{m}^2}{9M_\Delta^3F_0^2}\int_0^1{\mathrm{d}f_a}~
\left(f_am_{B0}-M_\Delta-m_{B0}\right)\\
\nn&\times\left\{7(M_\Delta(c_T + c_m) - 2 m_{B0} c_T)\lambda_1\left(\Delta_{\ref{Fdiagschptg}}\right)-(19M_\Delta(c_T + c_m) - 8 m_{B0} c_T)\rho_1\left(\Delta_{\ref{Fdiagschptg}}\right)\right\},\\[1em]
\nn F_{1,\ref{Fdiagschpti}}&=\frac{\mathcal{C}^2c_m\text{Is}_{m}^2}{M_\Delta^2F_0^2}\int_0^1{\mathrm{d}f_a}\int_0^{1-f_a}{\mathrm{d}f_b}~
\Bigg[\left(2m_{B0}^2-\frac{8q^2}{3}\right)\lambda_1\left(\Delta_{\ref{Fdiagschpti}}\right)-\left(m_{B0}^2-\frac{4q^2}{9}\right)\rho_1\left(\Delta_{\ref{Fdiagschpti}}\right)\\[1em]
\nn&+\Big(q^42f_a(f_a+f_b-1)+q^2m_{B0}\left[m_{B0}(16f_af_b-4f_a(f_a+4)+21f_b^2-34f_b+13)+M_\Delta(-20f_a-21f_b+13)\right]\\[1em]
\nn&+4m_{B0}^3(10f_a+8f_b-5)(-f_bm_{B0}+M_\Delta+m_{B0})\Big)\frac{\lambda_2\left(\Delta_{\ref{Fdiagschpti}}\right)}{3}\\[1em]
\nn&+(4m_{B0}^2q^2-q^4)\Big[-\frac{2}{3}f_am_{B0}(f_a+f_b-1)(2f_a+2f_b-1)(-f_bm_{B0}+M_\Delta+m_{B0})\Big]\lambda_3\left(\Delta_{\ref{Fdiagschpti}}\right)\\[1em]
\nn&+\Big(q^4(f_a+f_b-1)f_a+ q^22m_{B0}\left[m_{B0}(-20f_af_b+2f_a(f_a+10)-21f_b^2+32f_b-11)+M_\Delta(22f_a+21f_b-11)\right]\\[1em]
\nn&-4m_{B0}^3(22f_a+20f_b-11)(-f_bm_{B0}+M_\Delta+m_{B0})\Big)\frac{\rho_2\left(\Delta_{\ref{Fdiagschpti}}\right)}{18}\Bigg],\\[1em]
\nn F_{1,\ref{Fdiagschptj}}&=\frac{\mathcal{C}^2c_T\text{Is}_{m}^2}{9M_\Delta^4F_0^2}\int_0^1{\mathrm{d}f_a}\int_0^{1-f_a}{\mathrm{d}f_b}~
\Bigg[\Bigg(2(-2M_\Delta m_{B0}q^2(15f_a+3f_b-10)\\[1em]
\nn&+q^2(-2m_{B0}^2(10f_a+10f_b-9)+4q^2(f_a+f_b)-q^2)+6M_\Delta^2(3m_{B0}^2+2q^2))\Bigg)\lambda_1\left(\Delta_{\ref{Fdiagschptj}}\right)\\[1em]
\nn&+\Bigg(M_\Delta^2\bigg(-3m_{B0}^2q^2(4f_af_b+3f_a(7f_a-5)+f_b^2-3f_b+2)+6m_{B0}^4(13f_a-7f_b-3)(f_a+f_b-1)\\[1em]
\nn&+q^4(6f_af_b-3f_a-3f_b+1)\bigg)+M_\Delta m_{B0}q^2\bigg(m_{B0}^2(f_a+f_b-1)(18f_af_b+5f_a(3f_a-1)+3f_b^2-17f_b+6)\\[1em]
\nn&+q^2((5-3f_a)f_b^2+(16-15f_a)f_af_b+f_a(5f_a-11)-11f_b+2)\bigg)\\[1em]
\nn&+q^2\bigg(-m_{B0}^2q^2(f_a+f_b-1)(34f_af_b+f_a(2f_a-7)+2f_b^2-7f_b+1)+2m_{B0}^4(f_a+f_b-1)^2(5f_a+5f_b-1)\\[1em]
\nn&+f_af_bq^4(7f_a+7f_b-4)\bigg)+2M_\Delta^3m_{B0}(6m_{B0}^2(-8f_a+2f_b+3)+(24f_a-1)q^2)+18M_\Delta^4m_{B0}^2\Bigg)\lambda_2\left(\Delta_{\ref{Fdiagschptj}}\right)\\[1em]
\nn&+\Bigg(f_af_bq^2\bigg(M_\Delta m_{B0}q^2(q^2(f_a^2-3f_a+f_b^2-3f_b+2)-2m_{B0}^2(f_a+f_b-5)(f_a+f_b-1))\\[1em]
\nn&+M_\Delta^2(12m_{B0}^4(3f_a-f_b-1)(f_a+f_b-1)-4(3(f_a-1)f_a+2)m_{B0}^2q^2+q^4)\\[1em]
\nn&+q^2(f_a+f_b-1)(4m_{B0}^2-q^2)(m_{B0}^2(f_a+f_b-1)^2-f_af_bq^2)\\[1em]
\nn&-2M_\Delta^3m_{B0}(12(2f_a-1)m_{B0}^2-6f_aq^2+q^2)+12M_\Delta^4m_{B0}^2\bigg)\Bigg)\lambda_3\left(\Delta_{\ref{Fdiagschptj}}\right)\\[1em]
\nn&+\Bigg(q^4\left[-22(f_a+f_b)+1\right]+q^2\left[M_\Delta m_{B0}(249f_a+21f_b-206)+2m_{B0}^2(79f_a+79f_b-69)-132M_\Delta^2\right]\\[1em]
\nn&-162M_\Delta^2m_{B0}^2\Bigg)\frac{\rho_1\left(\Delta_{\ref{Fdiagschptj}}\right)}{6}\\[1em]
\nn&+\Bigg(q^6\left[-17f_a-17f_b+5\right]f_af_b+q^4\bigg[M_\Delta m_{B0}(f_a^2(51f_b-31)+f_a(f_b(3f_b-74)+43)+(43-31f_b)f_b-4)\\[1em]
\nn&+2m_{B0}^2(f_a+f_b-1)(2f_a^2+f_a(46f_b-7)+f_b(2f_b-7)+1)+M_\Delta^2(f_a(27-69f_b)+27f_b-2)\bigg]\\[1em]
\nn&-q^2\bigg[-6M_\Delta^2m_{B0}^2(27f_a^2+f_a(11f_b-23)+f_b(5f_b-17)+8)\\[1em]
\nn&+M_\Delta m_{B0}^3(f_a+f_b-1)(51f_a^2+f_a(54f_b-49)+f_b(3f_b-97)+30)\\[1em]
\nn&+M_\Delta^3m_{B0}(75f_a-9f_b+26)+8m_{B0}^4(f_a+f_b-1)^2(4f_a+4f_b+1)+36M_\Delta^4\bigg]\\[1em]
\nn&-6M_\Delta^2m_{B0}^2(M_\Delta-m_{B0}(f_a+f_b-1))(m_{B0}(-31f_a+13f_b+9)+9M_\Delta)\Bigg)\frac{\rho_2\left(\Delta_{\ref{Fdiagschptj}}\right)}{6}\Bigg],\label{EqF1amps}\\[1em]
F_{1,\ref{Fdiagsvec}}&= -\text{Is}_V\frac{g_vF_V}{m_V}\frac{q^2}{q^2-m^2_V},\\[1em]
\nn F_{2,\ref{Fdiagschpta}}&=F_{2,\ref{Fdiagschptb}}=F_{2,\ref{Fdiagschptc}}=F_{2,\ref{Fdiagschptd}}=F_{2,\ref{Fdiagschptg}}=0,\\[1em]
\nn F_{2,\ref{Fdiagschpttree}}&=c_{b23}(b_D-d_{102}q^2)+c_b(b_F-d_{101}q^2),\\[1em]
\nn F_{2,\ref{Fdiagschpte}}&=\frac{c_{DF}^2c_m}{4F_0^2}\int_0^1{\mathrm{d}f_a}\int_0^{1-f_a}{\mathrm{d}f_b}
~\Big[-2\left(2f_a(4f_b+5)+8f_b^2+5f_b-7\right)\lambda_2\left(\Delta_{\ref{Fdiagschpte}}\right)\\[1em]
\nn&+2m_{B0}^2(2f_a+2f_b-1)\left[q^2\left(f_af_b^2+f_a(f_a+1)f_b+2(f_a-1)f_a\right)+f_b^3m_{B0}^2\right]\lambda_3\left(\Delta_{\ref{Fdiagschpte}}\right)\\[1em]
\nn&+m_{B0}^2(f_b+2)(2f_a+2f_b-1)\rho_2\left(\Delta_{\ref{Fdiagschpte}}\right)\Big],\\[1em]
\nn F_{2,\ref{Fdiagschptf}}&=\frac{c_{DF}^2c_m}{4F_0^2}\int_0^1{\mathrm{d}f_a}\int_0^{1-f_a}{\mathrm{d}f_b}
~\Big[8(3f_a-1)m_{B0}^2\lambda_2\left(\Delta_{\ref{Fdiagschptf}}\right)\\[1em]
\nn&+8f_am_{B0}^2((f_a-1)f_bq^2-m_{B0}^2(f_a+f_b)^2)\lambda_3\left(\Delta_{\ref{Fdiagschptf}}\right)-4f_am_{B0}^2\rho_2\left(\Delta_{\ref{Fdiagschptf}}\right)\Big],\\[1em]
\nn F_{2,\ref{Fdiagschpth}}&=\frac{4\mathcal{C}^2m_{B0}\text{Is}_{m}^2}{9M_\Delta^3F_0^2}\int_0^1{\mathrm{d}f_a}~
\left(f_am_{B0}-M_\Delta-m_{B0}\right)\\
\nn&\times\left\{2(M_\Delta(c_T + c_m) - 2 m_{B0} c_T)\lambda_1\left(\Delta_{\ref{Fdiagschptg}}\right)-(5M_\Delta(c_T + c_m) - 4 m_{B0} c_T)\rho_1\left(\Delta_{\ref{Fdiagschptg}}\right)\right\},\\[1em]
\nn F_{2,\ref{Fdiagschpti}}&=\frac{\mathcal{C}^2c_m\text{Is}_{m}^2}{M_\Delta^2F_0^2}\int_0^1{\mathrm{d}f_a}\int_0^{1-f_a}{\mathrm{d}f_b}~
\Bigg[\frac{4m_{B0}^2}{3}\left(4\lambda_1\left(\Delta_{\ref{Fdiagschpti}}\right)+\frac13\rho_1\left(\Delta_{\ref{Fdiagschpti}}\right)\right)\\[1em]
\nn&+\Big(-q^24m_{B0}\left[m_{B0}(f_a^2+6f_a(f_b-1)+5f_b^2-8f_b+3)+M_\Delta(-5f_a-5f_b+3)\right]\\[1em]
\nn&-4m_{B0}^3(10f_a+7f_b-4)(-f_bm_{B0}+M_\Delta+m_{B0})\Big)\frac{\lambda_2\left(\Delta_{\ref{Fdiagschpti}}\right)}{3}\\[1em]
\nn&+(4m_{B0}^2q^2-q^4)\Big[\frac{2}{3}f_am_{B0}(f_a+f_b-1)(2f_a+2f_b-1)(-f_bm_{B0}+M_\Delta+m_{B0})\Big]\lambda_3\left(\Delta_{\ref{Fdiagschpti}}\right)\\[1em]
\nn&+\Big(q^22m_{B0}\left[m_{B0}(-9f_af_b+f_a(2f_a+9)-11f_b^2+17f_b-6)+M_\Delta(11f_a+11f_b-6)\right]\\[1em]
\nn&+2m_{B0}^3(22f_a+22f_b-13)(-f_bm_{B0}+M_\Delta+m_{B0})\Big)\frac{\rho_2\left(\Delta_{\ref{Fdiagschpti}}\right)}{9}\Bigg],\\[1em]
\nn F_{2,\ref{Fdiagschptj}}&=\frac{\mathcal{C}^2c_T\text{Is}_{m}^2}{9M_\Delta^4F_0^2}\int_0^1{\mathrm{d}f_a}\int_0^{1-f_a}{\mathrm{d}f_b}~
\Bigg[\Bigg(-8m_{B0}(-M_\Delta(m_{B0}^2(6f_a+6f_b-13))-4m_{B0}^3(f_a+f_b-1)+15M_\Delta^2m_{B0})\\[1em]
\nn&+q^28m_{B0}M_\Delta(6f_a+1)\Bigg)\lambda_1\left(\Delta_{\ref{Fdiagschptj}}\right)\\[1em]
\nn&+\Bigg(-4m_{B0}(M_\Delta(m_{B0}^2q^2(3f_a^3+4f_a^2+(f_a+9)(3f_a-1)f_b-12f_a+4f_b^2+1)\\[1em]
\nn&+3m_{B0}^4(f_a+f_b-2)(f_a+f_b-1)^2-f_a(3f_a+4)f_bq^4)+M_\Delta^3(m_{B0}^2(-24f_a+6f_b+10)+(12f_a-1)q^2)\\[1em]
\nn&+M_\Delta^2m_{B0}(15m_{B0}^2(f_a-f_b)(f_a+f_b-1)+q^2(3f_a(-5f_a+2f_b+2)-3f_b-1))+2m_{B0}^5(f_a+f_b-1)^3\\[1em]
\nn&-m_{B0}^3q^2(f_a+f_b-1)(22f_af_b-5f_a-5f_b+1)+f_af_bm_{B0}q^4(5f_a+5f_b-2)+6M_\Delta^4m_{B0})\Bigg)\lambda_2\left(\Delta_{\ref{Fdiagschptj}}\right)\\[1em]
\nn&+\Bigg(-4f_af_bm_{B0}q^2(M_\Delta(m_{B0}^2q^2(6f_af_b+f_a(2f_a-3)+2f_b^2-3f_b+1)-6m_{B0}^4(f_a+f_b-1)^2-f_af_bq^4)\\[1em]
\nn&+M_\Delta^2m_{B0}(3m_{B0}^2(3f_a-f_b-1)(f_a+f_b-1)+(-3(f_a-1)f_a-1)q^2)\\[1em]
\nn&+m_{B0}(f_a+f_b-1)(4m_{B0}^2-q^2)(m_{B0}^2(f_a+f_b-1)^2-f_af_bq^2)\\[1em]
\nn&-M_\Delta^3(4(3f_a-2)m_{B0}^2-3f_aq^2+q^2)+3M_\Delta^4m_{B0})\Bigg)\lambda_3\left(\Delta_{\ref{Fdiagschptj}}\right)\\[1em]
\nn&+\Bigg(m_{B0}(m_{B0}(M_\Delta m_{B0}(-21f_a-21f_b+82)+26m_{B0}^2(f_a+f_b-1)+138M_\Delta^2)\\[1em]
\nn&-q^2\left[24m_{B0}(f_a+f_b)+57f_aM_\Delta+M_\Delta-9m_{B0}\right])\Bigg)\frac{\rho_1\left(\Delta_{\ref{Fdiagschptj}}\right)}{3}\\[1em]
\nn&+\Bigg(m_{B0}(M_\Delta(m_{B0}^2q^2((9f_a+17)f_b^2+f_a(21f_a+58)f_b+f_a(f_a(12f_a+17)-39)-27f_b+2)\\[1em]
\nn&+3m_{B0}^4(f_a+f_b-10)(f_a+f_b-1)^2-f_a(12f_a+5)f_bq^4)\\[1em]
\nn&+2m_{B0}^3q^2(f_a+f_b-1)(-16f_af_b+f_a(3f_a+2)+3f_b^2+2f_b-1)\\[1em]
\nn&+M_\Delta^3(m_{B0}^2(-69f_a-3f_b+38)+(21f_a-2)q^2)\\[1em]
\nn&+M_\Delta^2m_{B0}(6m_{B0}^2(4f_a-7f_b+7)(f_a+f_b-1)-q^2(3f_a(11f_a-20f_b+4)+21f_b+2))\\[1em]
\nn&-8m_{B0}^5(f_a+f_b-1)^3+f_af_bm_{B0}q^4(7f_a+7f_b+5)+42M_\Delta^4m_{B0})\Bigg)\frac{\rho_2\left(\Delta_{\ref{Fdiagschptj}}\right)}{3}\Bigg],\\[1em]
F_{2,\ref{Fdiagsvec}}&= -\text{Is}_V\frac{g_tF_V}{m_V}\frac{q^2}{q^2-m^2_V}.\label{EqF2amps}
\end{align}

\begin{table}
\footnotesize
    \begin{tabular}{cccccccc}
\toprule
$p$ & $n$ & $\Sigma^+$ & $\Sigma^0$& $\Sigma^-$ & $\Lambda$ & $\Xi^0$ & $\Xi^-$\\\midrule
$\frac{1}{3}$&$-\frac{2}{3}$&$\frac{1}{3}$&$\frac{1}{3}$&$\frac{1}{3}$&$-\frac{1}{3}$&$-\frac{2}{3}$&$\frac{1}{3}$\\
\bottomrule
    \end{tabular}
  \caption{Values of the isospin constant $c_{b23}$ for the higher-order coupling of a photon to an octet baryon.}
  \label{tab:cb23}
\end{table}

\begin{table}
\footnotesize
    \begin{tabular}{l cccc}
\toprule
&$p$ & $n$ & $\Sigma^+$ & $\Sigma^0$\\
\midrule
$p$ &$D+F$ &$\sqrt{2}(D+F)$ &$\sqrt{2}(D-F)$ &$D-F$\\
$n$ &$\sqrt{2}(D+F)$ &$-(D+F)$ &$0$ &$F-D$\\
$\Sigma^+$ &$\sqrt{2}(D-F)$ & $0$& $2F$&$-2F$\\
$\Sigma^0$ & $D-F$& $F-D$&$-2F$ &$0$\\
$\Sigma^-$ & $0$& $\sqrt{2}(D-F)$& $0$&$2F$\\
$\Lambda$ & $-\frac{D+3F}{\sqrt{3}}$& $-\frac{D+3F}{\sqrt{3}}$& $\frac{2D}{\sqrt{3}}$&$\frac{2D}{\sqrt{3}}$\\
$\Xi^0$ & $0$& $0$&$\sqrt{2}(D+F)$ &$-(D+F)$\\
$\Xi^-$ & $0$&$0$ & $0$&$D+F$\\
\bottomrule\\
\toprule
& $\Sigma^-$ & $\Lambda$ & $\Xi^0$ & $\Xi^-$\\
\midrule
$p$  &$0$ &$-\left(\frac{D+3F}{\sqrt{3}}\right)$ &$0$ &$0$\\
$n$ &$\sqrt{2}(D-F)$ &$-\left(\frac{D+3F}{\sqrt{3}}\right)$ &$0$ &$0$\\
$\Sigma^+$ &$0$ &$\frac{2D}{\sqrt{3}}$ &$\sqrt{2}(D+F)$ &$0$\\
$\Sigma^0$ & $2F$& $\frac{2D}{\sqrt{3}}$& $-(D+F)$&$D+F$\\
$\Sigma^-$ & $-2F$& $\frac{2D}{\sqrt{3}}$&$0$ &$\sqrt{2}(D+F)$\\
$\Lambda$ & $\frac{2D}{\sqrt{3}}$& $0$& $\frac{3F-D}{\sqrt{3}}$&$\frac{3F-D}{\sqrt{3}}$ \\
$\Xi^0$ & $0$& $\frac{3F-D}{\sqrt{3}}$& $F-D$& $\sqrt{2}(D-F)$\\
$\Xi^-$ & $\sqrt{2}(D+F)$& $\frac{3F-D}{\sqrt{3}}$&$\sqrt{2}(D-F)$ &$D-F$ \\
\bottomrule
    \end{tabular}
  \caption{Values of the isospin constant $c_{DF}$ for the different channels of the octet-baryon-to-octet-baryon transition via a pion or a kaon.}
  \label{tab:cdf}
\end{table}

\begin{table}
\footnotesize
    \begin{tabular}{cccccccc}
\toprule
$p$ & $n$ & $\Sigma^+$ & $\Sigma^0$& $\Sigma^-$ & $\Lambda$ & $\Xi^0$ & $\Xi^-$\\
\midrule
$\frac{3F-D}{\sqrt{3}}$&$\frac{3F-D}{\sqrt{3}}$&$\frac{2D}{\sqrt{3}}$&$\frac{2D}{\sqrt{3}}$&$\frac{2D}{\sqrt{3}}$&$-\frac{2D}{\sqrt{3}}$&$-\frac{3F+D}{\sqrt{3}}$&$-\frac{3F+D}{\sqrt{3}}$\\
\bottomrule
    \end{tabular}
  \caption{Values of the isospin constant $c_{DF}$ for the coupling of an $\eta$ meson to an octet baryon.}
  \label{tab:cdf_eta}
\end{table}

\begin{table}
\footnotesize
    \begin{tabular}{cccccccc}
\toprule
$p$ & $n$ & $\Sigma^+$ & $\Sigma^0$& $\Sigma^-$ & $\Lambda$ & $\Xi^0$ & $\Xi^-$\\
\midrule
$-\frac{1}{4}$&$\frac{1}{4}$&$-\frac{1}{2}$&$0$&$\frac{1}{2}$&$0$&$-\frac{1}{4}$&$\frac{1}{4}$\\
\bottomrule
    \end{tabular}
  \caption{Values of the isospin constant Is$_{mm}$ for the coupling of two pions to an octet baryon.}
  \label{tab:IsmmPi}
\end{table}

\begin{table}
\footnotesize
    \begin{tabular}{cccccccc}
\toprule
$p$ & $n$ & $\Sigma^+$ & $\Sigma^0$& $\Sigma^-$ & $\Lambda$ & $\Xi^0$ & $\Xi^-$\\
\midrule
$-\frac{1}{2}$&$-\frac{1}{4}$&$-\frac{1}{4}$&$0$&$\frac{1}{4}$&$0$&$\frac{1}{4}$&$\frac{1}{2}$\\
\bottomrule
    \end{tabular}
  \caption{Values of the isospin constant Is$_{mm}$ for the coupling of two kaons to an octet baryon.}
  \label{tab:IsmmKaon}
\end{table}

\begin{table}
\footnotesize
    \begin{tabular}{l cccccccc}
\toprule
&$p$ & $n$ & $\Sigma^+$ & $\Sigma^0$& $\Sigma^-$ & $\Lambda$ & $\Xi^0$ & $\Xi^-$\\
\midrule
$\Delta^{++}$&$-1$&$0$&$1$&$0$&$0$&$0$&$0$&$0$\\
$\Delta^{+}$&$\sqrt{\frac{2}{3}}$&$-\frac{\sqrt{3}}{3}$&$\frac{\sqrt{3}}{3}$&$-\sqrt{\frac{2}{3}}$&$0$&$0$&$0$&$0$\\
$\Delta^{0}$&$\frac{\sqrt{3}}{3}$&$\sqrt{\frac{2}{3}}$&$0$&$-\sqrt{\frac{2}{3}}$&$-\frac{\sqrt{3}}{3}$&$0$&$0$&$0$\\
$\Delta^{-}$&$0$&$1$&$0$&$0$&$-1$&$0$&$0$&$0$\\
$\Sigma^{*+}$&$-\frac{\sqrt{3}}{3}$&$0$&$-\frac{\sqrt{6}}{6}$&$\frac{\sqrt{6}}{6}$&$0$&$\frac{\sqrt{2}}{2}$&$\frac{\sqrt{3}}{3}$&$0$\\
$\Sigma^{*0}$&$\frac{\sqrt{6}}{6}$&$-\frac{\sqrt{6}}{6}$&$-\frac{\sqrt{6}}{6}$&$0$&$\frac{\sqrt{6}}{6}$&$-\frac{\sqrt{2}}{2}$&$\frac{\sqrt{6}}{6}$&$-\frac{\sqrt{6}}{6}$\\
$\Sigma^{*-}$&$0$&$\frac{\sqrt{3}}{3}$&$0$&$\frac{\sqrt{6}}{6}$&$-\frac{\sqrt{6}}{6}$&$-\frac{\sqrt{2}}{2}$&$0$&$-\frac{\sqrt{3}}{3}$\\
$\Xi^{*0}$&$0$&$0$&$-\frac{\sqrt{3}}{3}$&$\frac{\sqrt{6}}{6}$&$0$&$\frac{\sqrt{2}}{2}$&$-\frac{\sqrt{6}}{6}$&$\frac{\sqrt{3}}{3}$\\
$\Xi^{*-}$&$0$&$0$&$0$&$\frac{\sqrt{6}}{6}$&$\frac{\sqrt{3}}{3}$&$-\frac{\sqrt{2}}{2}$&$-\frac{\sqrt{3}}{3}$&$-\frac{\sqrt{6}}{6}$\\
$\Omega^{-}$&$0$&$0$&$0$&$0$&$0$&$0$&$-1$&$1$\\
\bottomrule
    \end{tabular}
  \caption{Values of the isospin constant Is$_m$ for the different channels of the decuplet-to-octet baryon transition via a pion or a kaon.}
  \label{tab:Ism}
\end{table}

\begin{table}
\footnotesize
    \begin{tabular}{cccccccccc}
\toprule
$\Delta^{++}$ & $\Delta^{+}$ & $\Delta^{0}$ & $\Delta^{-}$& $\Sigma^{*+}$ & $\Sigma^{*0}$ & $\Sigma^{*-}$ & $\Xi^{*0}$ & $\Xi^{*-}$& $\Omega$\\
\midrule
$0$&$0$&$0$&$0$&$-\frac{\sqrt{2}}{2}$&$\frac{\sqrt{2}}{2}$&$\frac{\sqrt{2}}{2}$&$-\frac{\sqrt{2}}{2}$&$\frac{\sqrt{2}}{2}$&$0$\\
\bottomrule
    \end{tabular}
  \caption{Values of the isospin constant Is$_m$ for the decuplet-to-octet baryon transition via an $\eta$ meson.}
  \label{tab:Ism_eta}
\end{table}

For a check of our results, we compared them to those obtained in $SU(2)$ in the work of Ledwig et al.~\cite{Ledwig:2011cx} by setting to zero the kaon and $\eta$-meson contributions. We fully reproduce the analytical and numerical results, except for those of the diagram in \cref{Fdiagschptj}. For this particular diagram, in the analytical expression, the terms proportional to $q^4$ and higher were forgotten. This changed the numerical result, and here we correct this problem.

\section{Power-counting breaking terms}\label{SappPCBT}
In the EOMS scheme, the PCBTs are also absorbed into redefinitions of the LECs. The PCBTs for the diagrams with intermediate spin-1/2 baryon states vanish for the form factor $F_1$. In the particular renormalization scheme $\widetilde{\text{MS}}$ they vanish diagram by diagram, while for other schemes (e.g.\ $\overline{\text{MS}}$) they end up cancelling between diagrams. The only contributions to $F_2$ come from the diagrams in \cref{Fdiagschpte,Fdiagschptf}, and they have the following simple expressions:
\begin{align}
\nn F_{2,PCBT,\ref{Fdiagschpte}}&=\frac{c_{DF}^2c_mm_{B0}^2}{16\pi^2F_0^2},\\[1em]
 F_{2,PCBT,\ref{Fdiagschptf}}&=-\frac{c_{DF}^2c_bm_{B0}^2}{16\pi^2F_0^2}.
\end{align}
Concerning those diagrams with intermediate spin-3/2 baryon states, most of the contributions vanish as well. The only pieces that survive are as follows:
\begin{align}
\nn F_{1,PCBT,\ref{Fdiagschptg}}&=\frac{\mathcal{C}^2\text{Is}_m^2}{144\pi^2F_0^2}\Bigg\{
\frac{\left(M_\Delta^3+2 M_\Delta^2 m_{B0}-2 M_\Delta m_{B0}^2-6 m_{B0}^3\right)\left(M_\Delta(c_m+c_T)-2m_{B0}c_T\right)}{m_{B0}^4}M_\Delta^2\log\left[\frac{M_\Delta}{m_{B0}}\right]\\[1em]
\nn&-\frac{(M_\Delta-m_{B0})^3(M_\Delta+m_{B0})^5\left(M_\Delta(c_m+c_T)-2m_{B0}c_T\right)}{2M_\Delta^3m_{B0}^4}\log\left[\frac{M_\Delta^2-m_{B0}^2}{m_{B0}^2}\right]\\[1em]
\nn&+c_m\frac{-6 M_\Delta^6-12 M_\Delta^5 m_{B0}+9 M_\Delta^4 m_{B0}^2+36 M_\Delta^3 m_{B0}^3+10 M_\Delta^2 m_{B0}^4-16 M_\Delta m_{B0}^5-8 m_{B0}^6}{12M_\Delta^2 m_{B0}^2}\\[1em]
\nn&+c_T\frac{-6 M_\Delta^7+33 M_\Delta^5 m_{B0}^2+18 M_\Delta^4 m_{B0}^3+10 M_\Delta^3 m_{B0}^4-12 M_\Delta^2 m_{B0}^5+4 m_{B0}^7}{12M_\Delta^3 m_{B0}^2}
\Bigg\},\\[1em]
\nn F_{1,PCBT,\ref{Fdiagschpth}}&=\frac{\mathcal{C}^2\text{Is}_m^2}{144\pi^2F_0^2}\Bigg\{
-\frac{\left(M_\Delta^3+2 M_\Delta^2 m_{B0}-2 M_\Delta m_{B0}^2-6 m_{B0}^3\right)\left(7M_\Delta(c_m+c_T)-2m_{B0}c_T\right)}{m_{B0}^4}M_\Delta^2\log\left[\frac{M_\Delta}{m_{B0}}\right]\\[1em]
\nn&+\frac{(M_\Delta-m_{B0})^3(M_\Delta+m_{B0})^5\left(7M_\Delta(c_m+c_T)-2m_{B0}c_T\right)}{2M_\Delta^2m_{B0}^4}\log\left[\frac{M_\Delta^2-m_{B0}^2}{m_{B0}^2}\right]\\[1em]
\nn&+c_m\frac{42 M_\Delta^6+84 M_\Delta^5 m_{B0}-63 M_\Delta^4 m_{B0}^2-108 M_\Delta^3 m_{B0}^3-22 M_\Delta^2 m_{B0}^4+64M_\Delta m_{B0}^5+32 m_{B0}^6}{12M_\Delta^2 m_{B0}^2}\\[1em]
\nn&+c_T\frac{42 M_\Delta^7+72 M_\Delta^6 m_{B0}-87 M_\Delta^5 m_{B0}^2-90 M_\Delta^4 m_{B0}^3-22 M_\Delta^3 m_{B0}^4+60M_\Delta^2 m_{B0}^5+24M_\Delta m_{B0}^6-4 m_{B0}^7}{12M_\Delta^3 m_{B0}^2}
\Bigg\},\\[1em]
\nn F_{1,PCBT,\ref{Fdiagschpti}}&=\frac{\mathcal{C}^2\text{Is}_m^2c_m}{48\pi^2F_0^2}\Bigg\{
\nn\frac{5 M_\Delta^3+8 M_\Delta^2 m_{B0}-6 M_\Delta m_{B0}^2-12 m_{B0}^3}{m_{B0}^4}M_\Delta^3\log\left[\frac{M_\Delta}{m_{B0}}\right]\\[1em]
\nn&-\frac{(M_\Delta-m_{B0})^2(M_\Delta+m_{B0})^4(5 M_\Delta^2-2 M_\Delta m_{B0}+3 m_{B0}^2)}{2M_\Delta^2m_{B0}^4}\log\left[\frac{M_\Delta^2-m_{B0}^2}{m_{B0}^2}\right]\\[1em]
\nn&-\frac{30 M_\Delta^5+48 M_\Delta^4 m_{B0}-21 M_\Delta^3 m_{B0}^2-48 M_\Delta^2 m_{B0}^3-10 M_\Delta m_{B0}^4+8 m_{B0}^5}{12M_\Delta m_{B0}^2}
\Bigg\},\\[1em]
\nn F_{1,PCBT,\ref{Fdiagschptj}}&=\frac{\mathcal{C}^2\text{Is}_m^2c_T}{48\pi^2F_0^2}\Bigg\{
\frac{5 M_\Delta^3+8 M_\Delta^2 m_{B0}-6 M_\Delta m_{B0}^2-12 m_{B0}^3}{m_{B0}^4}M_\Delta^3\log\left[\frac{M_\Delta}{m_{B0}}\right]\\[1em]
\nn&-\frac{(M_\Delta-m_{B0})^2(M_\Delta+m_{B0})^4(5 M_\Delta^2-2 M_\Delta m_{B0}+3 m_{B0}^2)}{2M_\Delta^2m_{B0}^4}\log\left[\frac{M_\Delta^2-m_{B0}^2}{m_{B0}^2}\right]\\[1em]
&-\frac{30 M_\Delta^5+48 M_\Delta^4 m_{B0}-21 M_\Delta^3 m_{B0}^2-48 M_\Delta^2 m_{B0}^3-10 M_\Delta m_{B0}^4+8 m_{B0}^5}{12M_\Delta m_{B0}^2}
\Bigg\},\label{EqPCBT1amps}\\[1em]
\nn F_{2,PCBT,\ref{Fdiagschpth}}&=\frac{\mathcal{C}^2\text{Is}_m^2}{36\pi^2F_0^2}\Bigg\{
\frac{\left(M_\Delta^3+2 M_\Delta^2 m_{B0}-2 M_\Delta m_{B0}^2-6 m_{B0}^3\right)\left(2M_\Delta(c_m+c_T)-m_{B0}c_T\right)}{m_{B0}^4}M_\Delta^2\log\left[\frac{M_\Delta}{m_{B0}}\right]\\[1em]
\nn&-\frac{(M_\Delta-m_{B0})^3(M_\Delta+m_{B0})^5\left(2M_\Delta(c_m+c_T)-m_{B0}c_T\right)}{2M_\Delta^3m_{B0}^4\left(2M_\Delta(c_m+c_T)-m_{B0}c_T\right)}\log\left[\frac{M_\Delta^2-m_{B0}^2}{m_{B0}^2}\right]\\[1em]
\nn&-\frac{6 M_\Delta^6+12 M_\Delta^5 m_{B0}-9 M_\Delta^4 m_{B0}^2-18 M_\Delta^3 m_{B0}^3-4 M_\Delta^2 m_{B0}^4+10M_\Delta m_{B0}^5+5 m_{B0}^6}{6M_\Delta^2 m_{B0}^2}\\[1em]
\nn&+\frac{-12 M_\Delta^7-18 M_\Delta^6 m_{B0}+30 M_\Delta^5 m_{B0}^2+27 M_\Delta^4 m_{B0}^3+8 M_\Delta^3 m_{B0}^4-18M_\Delta^2 m_{B0}^5-6M_\Delta m_{B0}^6+2 m_{B0}^7}{12M_\Delta^2 m_{B0}^2}
\Bigg\},\\[1em]
\nn F_{2,PCBT,\ref{Fdiagschpti}}&=\frac{\mathcal{C}^2\text{Is}_m^2c_m}{36\pi^2F_0^2}\Bigg\{
-\frac{5 M_\Delta^3+4 M_\Delta^2 m_{B0}-10 M_\Delta m_{B0}^2-9 m_{B0}^3}{m_{B0}^4}M_\Delta^3\log\left[\frac{M_\Delta}{m_{B0}}\right]\\[1em]
\nn&+\frac{(M_\Delta+m_{B0})^3(5 M_\Delta^5-11 M_\Delta^4 m_{B0}+8 M_\Delta^3m_{B0}^2-5 M_\Delta^2m_{B0}^3+2 M_\Delta m_{B0}^4+m_{B0}^5)}{2M_\Delta^2m_{B0}^4}\log\left[\frac{M_\Delta^2-m_{B0}^2}{m_{B0}^2}\right]\\[1em]
\nn&+\frac{30 M_\Delta^6+24 M_\Delta^5 m_{B0}-45 M_\Delta^4 m_{B0}^2-42 M_\Delta^3 m_{B0}^3-44 M_\Delta^2 m_{B0}^4-16M_\Delta m_{B0}^5+7 m_{B0}^6}{12M_\Delta^2 m_{B0}^2}
\Bigg\},\\[1em]
\nn F_{2,PCBT,\ref{Fdiagschptj}}&=\frac{\mathcal{C}^2\text{Is}_m^2c_T}{108\pi^2F_0^2}\Bigg\{
-\frac{15 M_\Delta^5+13 M_\Delta^4 m_{B0}-47 M_\Delta^3 m_{B0}^2-34 M_\Delta^2 m_{B0}^3+62 M_\Delta m_{B0}^4+36 m_{B0}^5}{m_{B0}^4}M_\Delta\log\left[\frac{M_\Delta}{m_{B0}}\right]\\[1em]
\nn&+\frac{(M_\Delta+m_{B0})^4(M_\Delta-m_{B0})^2(15 M_\Delta^4-17 M_\Delta^3 m_{B0}+2 M_\Delta^2 m_{B0}^2+5 M_\Delta m_{B0}^3+m_{B0}^4)}{2M_\Delta^4m_{B0}^4}\log\left[\frac{M_\Delta^2-m_{B0}^2}{m_{B0}^2}\right]\\[1em]
\nn&+\frac{30 M_\Delta^8+26 M_\Delta^7 m_{B0}-79 M_\Delta^6 m_{B0}^2-55 M_\Delta^5 m_{B0}^3+45 M_\Delta^4 m_{B0}^4}{4M_\Delta^4 m_{B0}^2}\\[1em]
&+\frac{18 M_\Delta^3 m_{B0}^5-11 M_\Delta^2 m_{B0}^6-14 M_\Delta m_{B0}^7-6 m_{B0}^8}{4M_\Delta^4 m_{B0}^2}
\Bigg\}.\label{EqPCBT2amps}
\end{align}
\end{appendices}



\begin{thebibliography}{99}

\bibitem{Weinberg:1978kz}
 S.~Weinberg,
 \emph{Phenomenological Lagrangians},
 Physica A {\bf 96}, 327 (1979)
 [\href{http://www.sciencedirect.com/science/article/pii/0378437179902231?via\%3Dihub}{DOI:10.1016/0378-4371(79)90223-1}].

\bibitem{Gasser:1987rb} 
J.~Gasser, M.~E.~Sainio, and A.~Svarc, 
\emph{Nucleons with Chiral Loops}, 
Nucl.\ Phys.\ B {\bf 307}, 779 (1988) 
[\href{http://www.sciencedirect.com/science/article/pii/0550321388901083?via\%3Dihub}{DOI:10.1016/0550-3213(88)90108-3}].
 
\bibitem{Krause:1990xc}
  A.~Krause,
  \emph{Baryon Matrix Elements of the Vector Current in Chiral
                        Perturbation Theory},
  Helv.\ Phys.\ Acta {\bf 63}, 3 (1990)
  [\href{https://www.e-periodica.ch/cntmng?pid=hpa-001:1990:63::1083}{DOI:10.5169/seals-116214}].

\bibitem{Jenkins:1990jv}
 E.~E.~Jenkins and A.~V.~Manohar,
 \emph{Baryon chiral perturbation theory using a heavy fermion
                        Lagrangian},
 Phys.\ Lett.\ B {\bf 255}, 558 (1991)
 [\href{http://www.sciencedirect.com/science/article/pii/037026939190266S?via\%3Dihub}{DOI:10.1016/0370-2693(91)90266-S}].

\bibitem{Pascalutsa:2000kd}
 V.~Pascalutsa,
 \emph{Correspondence of consistent and inconsistent spin - 3/2
                        couplings via the equivalence theorem},
 Phys.\ Lett.\ B {\bf 503}, 85 (2001)
 [\href{https://arxiv.org/abs/hep-ph/0008026}{hep-ph/0008026}].

\bibitem{Pascalutsa:2002pi}
 V.~Pascalutsa and D.~R.~Phillips,
 \emph{Effective theory of the delta(1232) in Compton
                        scattering off the nucleon},
 Phys.\ Rev.\ C {\bf 67}, 055202 (2003)
 [\href{https://arxiv.org/abs/nucl-th/0212024}{nucl-th/0212024}

\bibitem{Pascalutsa:2006up}
 V.~Pascalutsa, M.~Vanderhaeghen, and S.-N.~Yang,
 \emph{Electromagnetic excitation of the
                        Delta(1232)-resonance},
 Phys.\ Rept.\ {\bf 437}, 125 (2007)
 [\href{https://arxiv.org/abs/hep-ph/0609004}{hep-ph/0609004}].

\bibitem{Borasoy:1995ds}
 B.~Borasoy and Ulf-G.~Mei{\ss}ner,
 \emph{Chiral Lagrangians for baryons coupled to massive spin 1
                        fields},
 Int.\ J.\ Mod.\ Phys. A {\bf 11}, 5183 (1996)
 [\href{https://arxiv.org/abs/hep-ph/9511320}{hep-ph/9511320}].

\bibitem{Drechsel:1998hk}
 D.~Drechsel, O.~Hanstein, S.~S.~Kamalov, and L.~Tiator,
 \emph{A Unitary isobar model for pion photoproduction and
                        electroproduction on the proton up to 1-GeV},
 Nucl.\ Phys.\ A {\bf 645}, 145 (1999)
 [\href{https://arxiv.org/abs/nucl-th/9807001}{nucl-th/9807001}].

\bibitem{Kubis:2000aa}
 B.~Kubis and Ulf-G.~Mei{\ss}ner,
 \emph{Baryon form-factors in chiral perturbation theory},
 Eur.\ Phys.\ J.\ C {\bf 18}, 747 (2001)
 [\href{https://arxiv.org/abs/hep-ph/0010283}{hep-ph/0010283}].
 
\bibitem{Kubis:2000zd}
B.~Kubis and Ulf-G.~Mei{\ss}ner, 
\emph{Low-energy analysis of the nucleon electromagnetic
                        form-factors}, 
Nucl.\ Phys.\ A {\bf 679}, 698 (2001) 
[\href{https://arxiv.org/abs/hep-ph/0007056}{hep-ph/0007056}].
 
\bibitem{Schindler:2005ke}
 M.~R.~Schindler, J.~Gegelia, and S.~Scherer,
 \emph{Electromagnetic form-factors of the nucleon in chiral
                        perturbation theory including vector mesons},
 Eur.\ Phys.\ J.\ A {\bf 26}, 1 (2005)
 [\href{https://arxiv.org/abs/nucl-th/0509005}{nucl-th/0509005}].
  
 \bibitem{Bauer:2012pv}
 T.~Bauer, J.~C.~Bernauer, and S.~Scherer,
 \emph{Electromagnetic form factors of the nucleon in effective
                        field theory},
 Phys.\ Rev.\ C {\bf 86}, 065206 (2012)
 [\href{https://arxiv.org/abs/1209.3872}{arXiv:1209.3872 [nucl-th]}].
 
\bibitem{Becher:1999he}
 T.~Becher and H.~Leutwyler,
 \emph{Baryon chiral perturbation theory in manifestly Lorentz
                        invariant form},
 Eur.\ Phys.\ J.\ C {\bf 9}, 643 (1999)
 [\href{https://arxiv.org/abs/hep-ph/9901384}{hep-ph/9901384}].
 
\bibitem{Becher:2001hv}
 T.~Becher and H.~Leutwyler,
 \emph{Low energy analysis of pi N $\rightarrow$ pi N},
 JHEP {\bf 06}, 017 (2001)
 [\href{https://arxiv.org/abs/hep-ph/0103263}{hep-ph/0103263}].
 
\bibitem{Scherer:2002tk}
      S.~Scherer,
      \emph{Introduction to chiral perturbation theory},
      Adv.\ Nucl.\ Phys.\ {\bf 27}, 277 (2003)
      [\href{https://arxiv.org/abs/hep-ph/0210398}{hep-ph/0210398}].
 
\bibitem{Ando:2006xy}
 S.-i.~Ando and H.~W.~Fearing,
 \emph{Ordinary muon capture on a proton in manifestly Lorentz invariant baryon chiral perturbation theory},
 Phys.\ Ref.\ D {\bf 75}, 014025 (2007)
 [\href{https://arxiv.org/abs/hep-ph/0608195}{hep-ph/0608195}].

\bibitem{Tang:1996ca}
 H.-B.~Tang,
 \emph{A New approach to chiral perturbation theory for matter
                        fields},
 (1996)
 [\href{https://arxiv.org/abs/hep-ph/9607436}{hep-ph/9607436}].

\bibitem{Ellis:1997kc}
 P.~J.~Ellis and H.-B.~Tang,
 \emph{Pion nucleon scattering in a new approach to chiral
                        perturbation theory},
 Phys.\ Rev.\ C {\bf 57}, 3356 (1998)
 [\href{https://arxiv.org/abs/hep-ph/9709354}{hep-ph/9709354}].

\bibitem{Gegelia:1999gf}
J.~Gegelia and G.~Japaridze, 
 \emph{Matching heavy particle approach to relativistic
                        theory},
 Phys.\ Rev.\ D {\bf 60}, 114038 (1999)
 [\href{https://arxiv.org/abs/hep-ph/9908377}{hep-ph/9908377}].

\bibitem{Fuchs:2003qc}
 T.~Fuchs, J.~Gegelia, G.~Japaridze, and S.~Scherer,
 \emph{Renormalization of relativistic baryon chiral
                        perturbation theory and power counting},
 Phys.\ Rev.\ D {\bf 68}, 056005 (2003)
 [\href{https://arxiv.org/abs/hep-ph/0302117}{hep-ph/0302117}].

\bibitem{Fuchs:2003ir}
T.~Fuchs, J.~Gegelia, and S.~Scherer, 
\emph{Electromagnetic form-factors of the nucleon in
                        relativistic baryon chiral perturbation theory}, 
J.\ Phys.\ G {\bf 30}, 1407 (2004) 
[\href{https://arxiv.org/abs/nucl-th/0305070}{nucl-th/0305070}].

\bibitem{Lehnhart:2004vi}
B.~C.~Lehnhart, J.~Gegelia, and S.~Scherer,
 \emph{Baryon masses and nucleon sigma terms in manifestly
                        Lorentz-invariant baryon chiral perturbation theory},
 J.\ Phys.\ G {\bf 31}, 89 (2005)
 [\href{https://arxiv.org/abs/hep-ph/0412092}{hep-ph/0412092}].

\bibitem{Schindler:2006ha}
 M.~R.~Schindler, D.~Djukanovic, J.~Gegelia, and S.~Scherer,
 \emph{Chiral expansion of the nucleon mass to order(q**6)},
 Phys.\ Lett.\ B {\bf 649}, 390 (2007)
 [\href{https://arxiv.org/abs/hep-ph/0612164}{hep-ph/0612164}].

\bibitem{Schindler:2006it}
M.~R.~Schindler, T.~Fuchs, J.~Gegelia, and S.~Scherer,
 \emph{Axial, induced pseudoscalar, and pion-nucleon
                        form-factors in manifestly Lorentz-invariant chiral
                        perturbation theory},
 Phys.\ Rev.\ C {\bf 75}, 025202 (2007)
 [\href{https://arxiv.org/abs/nucl-th/0611083}{nucl-th/0611083}].

\bibitem{Geng:2008mf}
L.~S.~Geng, J.~Martin Camalich, L.~Alvarez-Ruso, and M.~J.~Vicente Vacas, 
\emph{Leading SU(3)-breaking corrections to the baryon
                        magnetic moments in Chiral Perturbation Theory}, 
Phys.\ Rev.\ Lett.\ {\bf 101}, 222002 (2008) 
[\href{https://arxiv.org/abs/0805.1419}{arXiv:0805.1419 [hep-ph]}].

\bibitem{Geng:2009ik}
L.~S.~Geng, J.~Martin Camalich, and M.~J.~Vicente Vacas,
 \emph{SU(3)-breaking corrections to the hyperon vector
                        coupling f(1)(0) in covariant baryon chiral perturbation
                        theory},
 Phys.\ Rev.\ D {\bf 79}, 094022 (2009)
 [\href{https://arxiv.org/abs/0903.4869}{arXiv:0903.4869 [hep-ph]}].

\bibitem{MartinCamalich:2010fp}
 J.~Martin Camalich, L.~S.~Geng, and M.~J.~Vicente Vacas,
 \emph{The lowest-lying baryon masses in covariant SU(3)-flavor
                        chiral perturbation theory},
 Phys.\ Rev.\ D {\bf 82}, 074504 (2010)
 [\href{https://arxiv.org/abs/1003.1929}{arXiv:1003.1929 [hep-lat]}].

\bibitem{Ledwig:2011cx}
T.~Ledwig, J.~Martin Camalich, V.~Pascalutsa, and M.~Vanderhaeghen, 
\emph{The Nucleon and $\Delta$(1232) form factors at low
                        momentum-transfer and small pion masses}, 
Phys.\ Rev.\ D {\bf 85}, 034013 (2012) 
[\href{https://arxiv.org/abs/1108.2523}{arXiv:1108.2523 [hep-ph]}].

\bibitem{Alarcon:2011zs}
 J.~M.~Alarcon, J.~Martin Camalich, and J.~A.~Oller,
 \emph{The chiral representation of the $\pi N$ scattering
                        amplitude and the pion-nucleon sigma term},
 Phys.\ Rev.\ D {\bf 85}, 051503 (2012)
 [\href{https://arxiv.org/abs/1110.3797}{arXiv:1110.3797 [hep-ph]}].

\bibitem{Chen:2012nx}
Y.-H.~Chen, D.L.~Yao, and H.~Q.~Zheng,
 \emph{Analyses of pion-nucleon elastic scattering amplitudes
                        up to $O(p^4)$ in extended-on-mass-shell subtraction
                        scheme},
 Phys.\ Rev.\ D {\bf 87}, 054019 (2013)
 [\href{https://arxiv.org/abs/1212.1893}{arXiv:1212.1893 [hep-ph]}].

\bibitem{Alvarez-Ruso:2013fza}
 L.~Alvarez-Ruso, T.~Ledwig, J.~Martin Camalich, and M.~J.~Vicente Vacas,
 \emph{Nucleon mass and pion-nucleon sigma term from a chiral
                        analysis of lattice QCD data},
 Phys.\ Rev.\ D {\bf 88}, 054507 (2013)
 [\href{https://arxiv.org/abs/1304.0483}{arXiv:1304.0483 [hep-ph]}].

\bibitem{Ledwig:2014rfa}
T.~Ledwig, J.~Martin Camalich, L.~S.~Geng, and M.~J.~Vicente Vacas, 
\emph{Octet-baryon axial-vector charges and SU(3)-breaking effects in the semileptonic hyperon decays},
Phys.\ Rev.\ D {\bf 90}, 054502 (2014) 
[\href{https://arxiv.org/abs/1405.5456}{arXiv:1405.5456 [hep-ph]}].

\bibitem{Lensky:2014dda}
 V.~Lensky, J.~M.~Alarcon, and V.~Pascalutsa,
 \emph{Moments of nucleon structure functions at
                        next-to-leading order in baryon chiral perturbation
                        theory},
 Phys.\ Rev.\ C {\bf 90}, 055202 (2014)
 [\href{https://arxiv.org/abs/1407.2574}{arXiv:1407.2574 [hep-ph]}].

\bibitem{Blin:2016itn}
 A.~N.~Hiller Blin, T.~Ledwig, and M.~J.~Vicente Vacas,
 \emph{$\Delta(1232)$ resonance in the
                        $\vec{\gamma}p\rightarrow p\pi^0$ reaction at threshold},
 Phys.\ Lett.\ B {\bf 747}, 217 (2015)
 [\href{https://arxiv.org/abs/1412.4083}{arXiv:1412.4083 [hep-ph]}].

\bibitem{Blin:2014rpa}
 A.~N.~Hiller Blin, T.~Ledwig, and M.~J.~Vicente Vacas,
 \emph{Chiral dynamics in the $\vec{\gamma}p \to p\pi^0$
                        reaction},
 Phys.\ Rev.\ D {\bf 93}, 094018 (2016)
 [\href{https://arxiv.org/abs/1602.08967}{arXiv:1602.08967 [hep-ph]}].

\bibitem{HillerBlin:2016jpb}
A.~N.~Hiller Blin, 
\emph{Electromagnetic interactions of light baryons in covariant chiral perturbation theory},
PhD thesis at Universidad de Valencia (2016) 
[\href{http://roderic.uv.es/bitstream/handle/10550/56957/Tesis.pdf?sequence=1\&isAllowed=y}{http://roderic.uv.es/bitstream/handle/10550/56957/Tesis.pdf?sequence=1\&isAllowed=y}].
      
      \bibitem{Schindler:2003xv}
      M.~R.~Schindler, J.~Gegelia, and S.~Scherer,
      \emph{Infrared regularization of baryon chiral perturbation
                        theory reformulated},
      Phys.\ Lett.\ B {\bf 586}, 258 (2004)
      [\href{https://arxiv.org/abs/hep-ph/0309005}{hep-ph/0309005].}
 
 \bibitem{Scherer:2012xha}
      S.~Scherer and M.~R.~Schindler,
      \emph{A Primer for Chiral Perturbation Theory},
      Lect.\ Notes Phys.\ {\bf 830}, 1 (2012)
      [\href{https://link.springer.com/book/10.1007\%2F978-3-642-19254-8}{DOI:10.1007/978-3-642-19254-8}].

\bibitem{Siemens:2016jwj}
  D.~Siemens, J.~Ruiz de Elvira, E.~Epelbaum, M.~Hoferichter, H.~Krebs, B.~Kubis, and Ulf-G.~Mei{\ss}ner,
  \emph{Reconciling threshold and subthreshold expansions for
                        pion--nucleon scattering},
  Phys.\ Lett.\ B {\bf 770}, 27 (2017)
  [\href{https://arxiv.org/abs/1610.08978}{arXiv:1610.08978}].

\bibitem{Perdrisat:2006hj}
C.~F.~Perdrisat, V.~Punjabi, M.~Vanderhaeghen, 
\emph{Nucleon Electromagnetic Form Factors},
Prog.\ Part.\ Nucl.\ Phys.\ {\bf 59}, 694 (2007)
[\href{https://arxiv.org/abs/hep-ph/0612014}{hep-ph/0612014}].

\bibitem{Punjabi:2015bba}
V.~Punjabi, C.~F.~Perdrisat, M.~K.~Jones, E.~J.~Brash, and C.~E.~Carlson, 
\emph{The Structure of the Nucleon: Elastic Electromagnetic Form Factors}, 
Eur.\ Phys.\ J.\ A {\bf 51}, 79 (2015)
[\href{https://arxiv.org/abs/1503.01452}{arXiv:1503.01452 [nucl-ex]}].

\bibitem{Hofstadter:1955ae}
R.~Hofstadter and R.~W.~McAllister, 
\emph{Electron Scattering From the Proton}, 
Phys.\ Rev.\ {\bf 98}, 217 (1955) 
[\href{http://journals.aps.org/pr/abstract/10.1103/PhysRev.98.217}{DOI:10.1103/PhysRev.98.217}].

\bibitem{Hofstadter:1956qs}
R.~Hofstadter, 
\emph{Electron scattering and nuclear structure}, 
Rev.\ Mod.\ Phys.\ {\bf 28}, 214 (1956) 
[\href{http://journals.aps.org/rmp/abstract/10.1103/RevModPhys.28.214}{10.1103/RevModPhys.28.214}].

\bibitem{Yearian:1958shh}
M.~R.~Yearian and R.~Hofstadter, 
\emph{Magnetic Form Factor of the Neutron}, 
Phys.\ Rev.\ {\bf 110}, 552 (1958) 
[\href{http://journals.aps.org/pr/abstract/10.1103/PhysRev.110.552}{10.1103/PhysRev.110.552}].

\bibitem{Hofstadter:1959nya}
R.~Hofstadter, 
\emph{On nucleon structure}, 
Proceedings of ICHEP59, 355 (1960) 
[\href{https://inspirehep.net/record/1280981/files/c59-07-15-p355.pdf}{https://inspirehep.net/record/1280981/files/c59-07-15-p355.pdf}].

\bibitem{Bernauer:2010wm}
J.~C.~Bernauer et al. [A1 collaboration], 
\emph{High-precision determination of the electric and magnetic form factors of the proton}, 
Phys.\ Rev.\ Lett.\ {\bf 105}, 242001 (2010) 
[\href{https://arxiv.org/abs/1007.5076}{arXiv:1007.5076 [nucl-ex]}].

\bibitem{Pohl:2010zza}
R.~Pohl et al., 
\emph{The size of the proton}, 
Nature {\bf 466}, 213 (2010) 
[\href{http://www.nature.com/nature/journal/v466/n7303/full/nature09250.html}{DOI:10.1038/nature09250}].

\bibitem{Jentschura:2010ha}
U.~D.~Jentschura, 
\emph{Lamb Shift in Muonic Hydrogen. II. Analysis of the
                        Discrepancy of Theory and Experiment}, 
Annals Phys.\ {\bf 326}, 516 (2011) 
[\href{https://arxiv.org/abs/1011.5453}{arXiv:1011.5453 [hep-ph]}].

\bibitem{Miller:2011yw}
G.~A.~Miller, A.~W.~Thomas, J.~D.~Carroll, and J.~Rafelski, 
\emph{Natural Resolution of the Proton Size Puzzle}, 
Phys.\ Rev.\ A {\bf 84}, 020101 (2011) 
[\href{https://arxiv.org/abs/1101.4073}{arXiv:1101.4073 [physics.atom-ph]}].

\bibitem{Alarcon:2013cba} 
 J.~M.~Alarcon, V.~Lensky, and V.~Pascalutsa,
 \emph{Chiral perturbation theory of muonic hydrogen Lamb shift: polarizability contribution},
 Eur.\ Phys.\ J.\ C {\bf 74}, 2852 (2014)
 [\href{https://arxiv.org/abs/1312.1219}{arXiv:1312.1219 [hep-ph]}].

\bibitem{Carlson:2011zd}
C.~E.~Carlson and M.~Vanderhaeghen, 
\emph{Higher order proton structure corrections to the Lamb
                        shift in muonic hydrogen}, 
Phys.\ Rev.\ A {\bf 84}, 020102 (2011) 
[\href{https://arxiv.org/abs/arXiv:1101.5965}{arXiv:1101.5965 [hep-ph]}].

\bibitem{Distler:2010zq}
M.~O.~Distler, J.~C.~Bernauer, and T.~Walcher, 
\emph{The RMS Charge Radius of the Proton and Zemach Moments}, 
Phys.\ Lett.\ B {\bf 696}, 343 (2011) 
[\href{https://arxiv.org/abs/1011.1861}{arXiv:1011.1861 [nucl-th]}].

\bibitem{Griffioen:2015hta}
K.~Griffioen, C.~Carlson, and S.~Maddox, 
\emph{Consistency of electron scattering data with a small
                        proton radius}, 
Phys.\ Rev.\ C {\bf 93}, 065207 (2016) 
[\href{https://arxiv.org/abs/1509.06676}{arXiv:1509.06676 [nucl-ex]}].

\bibitem{Mergell:1995bf}
 P.~Mergell, Ulf-G.~Mei{\ss}ner, and D.~Drechsel,
 \emph{Dispersion theoretical analysis of the nucleon
                        electromagnetic form-factors},
 Nucl.\ Phys.\ A {\bf 596}, 367 (1996)
 [\href{https://arxiv.org/abs/hep-ph/9506375}{hep-ph/9506375}].

\bibitem{Hammer:2003ai}
H.~W.~Hammer and Ulf-G.~Mei{\ss}ner, 
\emph{Updated dispersion theoretical analysis of the nucleon
                        electromagnetic form-factors}, 
Eur.\ Phys.\ J.\ A {\bf 20}, 469 (2004)
[\href{https://arxiv.org/abs/hep-ph/0312081}{hep-ph/0312081}].

\bibitem{Belushkin:2005ds}
M.~A.~Belushkin, H.~W.~Hammer, and Ulf-G.~Mei{\ss}ner, 
\emph{Novel evaluation of the two-pion contribution to the
                        nucleon isovector form-factors}, 
Phys.\ Lett.\ B {\bf 633}, 507 (2006) 
[\href{https://arxiv.org/abs/hep-ph/0510382}{hep-ph/0510382}].

\bibitem{Belushkin:2006qa}
 M.~A.~Belushkin, H.-W.~Hammer, and Ulf-G.~Mei{\ss}ner,
 \emph{Dispersion analysis of the nucleon form-factors
                        including meson continua},
 Phys.\ Rev.\ C {\bf 75}, 035202 (2007)
 [\href{https://arxiv.org/abs/hep-ph/0608337}{hep-ph/0608337}].

\bibitem{Silva:2005vp}
 A.~Silva, D.~Urbano, and K.~Goeke,
 \emph{Baryon form factors in the chiral quark-soliton model},
 Nucl.\ Phys.\ A {\bf 755}, 290 (2005)
 [\href{http://www.sciencedirect.com/science/article/pii/S0375947405004227?via\%3Dihub}{DOI:10.1016/j.nuclphysa.2005.03.030}].

\bibitem{Ramalho:2011pp}
 G.~Ramalho and K.~Tsushima,
 \emph{Octet baryon electromagnetic form factors in a
                        relativistic quark model},
 Phys.\ Rev.\ D {\bf 84}, (054014) 2011
 [\href{https://arxiv.org/abs/1107.1791}{arXiv:1107.1791 [hep-ph]}].

\bibitem{Liu:2013fda}
 X.~Y.~Liu, K.~Khosonthongkee, A.~Limphirat, and Y.~Yan,
 \emph{Study of baryon octet electromagnetic form factors in
                        perturbative chiral quark model},
 J.\ Phys.\ G {\bf 41}, 055008 (2014)
 [\href{https://arxiv.org/abs/1309.2063}{arXiv:1309.2063 [hep-ph]}].

\bibitem{Yamazaki:2009zq}
T.~Yamazaki, Y.~Aoki, T.~Blum, H.-W.~Lin, S.~Ohta, S.~Sasaki, R.~Tweedie, and J.~Zanotti, 
\emph{Nucleon form factors with 2+1 flavor dynamical
                        domain-wall fermions}, 
Phys.\ Rev.\ D {\bf 79}, 114505 (2009) 
[\href{https://arxiv.org/abs/0904.2039}{arXiv:0904.2039 [hep-lat]}].

\bibitem{Syritsyn:2009mx}
S.~N.~Syritsyn et al., 
\emph{Nucleon Electromagnetic Form Factors from Lattice QCD
                        using 2+1 Flavor Domain Wall Fermions on Fine Lattices and
                        Chiral Perturbation Theory}, 
Phys.\ Rev.\ D {\bf 81}, 034507 (2010) 
[\href{https://arxiv.org/abs/0907.4194}{arXiv:0907.4194 [hep-lat]}].

\bibitem{Bratt:2010jn}
J.~D.~Bratt et al. [LHPC collaboration], 
\emph{Nucleon structure from mixed action calculations using
                        2+1 flavors of asqtad sea and domain wall valence
                        fermions}, 
Phys.\ Rev.\ D {\bf 82}, 094502 (2010) 
[\href{https://arxiv.org/abs/1001.3620}{arXiv:1001.3620 [hep-lat]}].

\bibitem{Alexandrou:2011db}
C.~Alexandrou, M.~Brinet, J.~Carbonell, M.~Constantinou, P.~A.~Harraud, P.~Guichon, K.~Jansen, T.~Korzec, and M.~Papinutto, 
\emph{Nucleon electromagnetic form factors in twisted mass
                        lattice QCD}, 
Phys.\ Rev.\ D {\bf 83}, 094502 (2011) 
[\href{https://arxiv.org/abs/1102.2208}{arXiv:1102.2208 [hep-lat]}].

\bibitem{Collins:2011mk}
S.~Collins et al., 
\emph{Dirac and Pauli form factors from lattice QCD}, 
Phys.\ Rev.\ D {\bf 84}, 074507 (2011) 
[\href{https://arxiv.org/abs/1106.3580}{arXiv:1106.3580 [hep-lat]}].

\bibitem{Flores-Mendieta:2015wir}
 R.~Flores-Mendieta and M.~A.~Rivera-Ruiz,
 \emph{Dirac form factors and electric charge radii of baryons
                        in the combined chiral and 1/N$_c$ expansions},
 Phys.\ Rev.\ D {\bf 92}, 094026 (2015)
 [\href{https://arxiv.org/abs/1511.02932}{arXiv:1511.02932 [hep-ph]}.

\bibitem{Carrillo-Serrano:2016igi}
 M.~E.~Carrillo-Serrano, W.~Bentz, I.~C.~Clo\"et, A.~W.~Thomas,
 \emph{Baryon Octet Electromagnetic Form Factors in a confining
                        NJL model},
 Phys.\ Lett.\ B {\bf 759}, 178 (2016)
 [\href{https://arxiv.org/abs/1603.02741}{arXiv:1603.02741 [nucl-th]}.

\bibitem{Aliev:2013jda} 
  T.~M.~Aliev, K.~Azizi, and M.~Savci,
  \emph{Electromagnetic form factors of octet baryons in QCD},
  Phys.\ Lett.\ B {\bf 723}, 145 (2013)
  [\href{https://arxiv.org/abs/1303.6798}{arXiv:1303.6798 [hep-ph]}].

\bibitem{Bernard:1998gv}
V.~Bernard, H.~W.~Fearing, T.~R.~Hemmert, and Ulf-G.~Mei{\ss}ner, 
\emph{The form-factors of the nucleon at small momentum
                        transfer}, 
Nucl.\ Phys.\ A {\bf 635}, 121 (1998) [Erratum: Nucl.\ Phys.\ A {\bf 642}, 563 (1998)] 
[\href{https://arxiv.org/abs/hep-ph/9801297}{hep-ph/9801297}].

\bibitem{Kaiser:2003qp}
N.~Kaiser, 
\emph{Spectral functions of isoscalar scalar and isovector
                        electromagnetic form-factors of the nucleon at two loop
                        order}, 
Phys.\ Rev.\ C {\bf 68}, 025202 (2003) 
[\href{https://arxiv.org/abs/nucl-th/0302072}{nucl-th/0302072}].

\bibitem{Jiang:2009jn}
F.-J.~Jiang and B.~C.~Tiburzi, 
\emph{Hyperon Electromagnetic Properties in Two-Flavor Chiral
                        Perturbation Theory}, 
Phys.\ Rev.\ D {\bf 81}, 034017 (2010) 
[\href{https://arxiv.org/abs/0912.2077}{arXiv:0912.2077 [nucl-th]}].

\bibitem{Kubis:1999xb}
 B.~Kubis, T.~R.~Hemmert, and Ulf-G.~Mei{\ss}ner,
 \emph{Baryon form-factors},
 Phys.\ Lett.\ B {\bf 456}, 240 (1999)
 [\href{https://arxiv.org/abs/hep-ph/9903285}{hep-ph/9903285}].

\bibitem{Geng:2009hh}
L.~S.~Geng, J.~Martin Camalich, and M.~J.~Vicente Vacas, 
\emph{Leading-order decuplet contributions to the baryon
                        magnetic moments in Chiral Perturbation Theory}, 
Phys.\ Lett.\ B {\bf 676}, 63 (2009) 
[\href{https://arxiv.org/abs/0903.0779}{arXiv:0903.0779 [hep-ph]}].

\bibitem{Shanahan:2014uka} 
  P.~E.~Shanahan et al. [CSSM and QCDSF/UKQCD Collaborations],
  \emph{Magnetic form factors of the octet baryons from lattice QCD and chiral extrapolation},
  Phys.\ Rev.\ D {\bf 89}, 074511 (2014)
  [\href{https://arxiv.org/abs/1401.5862}{arXiv:1401.5862 [hep-lat]}].

\bibitem{Shanahan:2014cga} 
  P.~E.~Shanahan et al. [CSSM and QCDSF/UKQCD Collaborations],
  \emph{Electric form factors of the octet baryons from lattice QCD and chiral extrapolation},
  Phys.\ Rev.\ D {\bf 90}, 034502 (2014)
  [\href{https://arxiv.org/abs/1403.1965}{arXiv:1403.1965 [hep-lat]}].

\bibitem{Gasser:2015dwa}
 J.~Gasser, M.~Hoferichter, H.~Leutwyler, and A.~Rusetsky,
 \emph{Cottingham formula and nucleon polarisabilities},
 Eur.\ Phys.\ J.\ C {\bf 75}, 375 (2015)
 [\href{https://arxiv.org/abs/1506.06747}{arXiv:1506.06747 [hep-ph]}].

\bibitem{Strikman:2010pu}
 M.~Strikman and C.~Weiss,
 \emph{Quantifying the nucleon's pion cloud with transverse
                        charge densities},
 Phys.\ Rev.\ C {\bf 82}, 042201 (2010)
 [\href{https://arxiv.org/abs/1004.3535}{arXiv:1004.3535 [hep-ph]}].

\bibitem{Granados:2013moa}
 C.~Granados and C.~Weiss,
 \emph{Chiral dynamics and peripheral transverse densities},
 JHEP {\bf 01}, 092 (2014)
 [\href{https://arxiv.org/abs/arXiv:1308.1634}{arXiv:1308.1634 [hep-ph]}].

\bibitem{Alarcon:2017lkk}
 J.~M.~Alarc\'on, A.~N.~Hiller Blin, and C.~Weiss,
 \emph{Transverse densities of octet baryons from chiral
                        effective field theory},
 Few Body Syst.\ {\bf 58}, 121 (2017)
 [\href{https://arxiv.org/abs/1701.05871}{arXiv:1701.05871 [hep-ph]}].

\bibitem{Alarcon:2017asr}
 J.~M.~Alarc\'on, A.~N.~Hiller Blin, M.~J.~Vicente Vacas, and C.~Weiss,
 \emph{Peripheral transverse densities of the baryon octet from
                        chiral effective field theory and dispersion analysis},
 Nucl.\ Phys.\ A {\bf 964}, 181 (2017)
 [\href{https://arxiv.org/abs/1703.04534}{arXiv:1703.04534 [hep-ph]}].

\bibitem{Granados:2017cib}
 C.~Granados, S.~Leupold, and E.~Perotti,
 \emph{The electromagnetic Sigma-to-Lambda hyperon transition form factors at low energies},
 Eur.\ Phys.\ J.\ A {\bf 53}, 117 (2017)
 [\href{https://arxiv.org/abs/1701.09130}{arXiv:1701.09130 [hep-ph]}].

\bibitem{Chiang:2001as}
 W.-T.~Chiang, S.-N.~Yang, L.~Tiator, and D.~Drechsel,
 \emph{An Isobar model for eta photoproduction and
                        electroproduction on the nucleon},
 Nucl.\ Phys.\ A {\bf 700}, 429 (2002)
 [\href{https://arxiv.org/abs/nucl-th/0110034}{nucl-th/0110034}].

\bibitem{Machleidt:1987hj}
 R.~Machleidt, K.~Holinde, C.~Elster,
 \emph{The Bonn Meson Exchange Model for the Nucleon Nucleon
                        Interaction}, 
 Phys.\ Rept.\ {\bf 149}, 1 (1987)
 [\href{http://www.sciencedirect.com/science/article/pii/S0370157387800029?via\%3Dihub}{DOI:10.1016/S0370-1573(87)80002-9}].

\bibitem{Machleidt:2000ge}
 R.~Machleidt,
 \emph{The High precision, charge dependent Bonn
                        nucleon-nucleon potential (CD-Bonn)},
 Phys.\ Rev.\ C {\bf 63}, 024001 (2001)
 [\href{https://arxiv.org/abs/nucl-th/0006014}{nucl-th/0006014}].

\bibitem{Bincer:1959tz}
 A.~M.~Bincer,
 \emph{Electromagnetic structure of the nucleon},
 Phys.\ Rev.\ {\bf 118}, 855 (1960)
 [\href{https://journals.aps.org/pr/abstract/10.1103/PhysRev.118.855}{DOI:10.1103/PhysRev.118.855}].

\bibitem{Koch:2001ii}
 J.~H.~Koch, V.~Pascalutsa, and S.~Scherer,
 \emph{Hadron structure and the limitations of phenomenological
                        models in electromagnetic reactions},
 Phys.\ Rev.\ C {\bf 65}, 045202 (2002)
 [\href{https://arxiv.org/abs/nucl-th/0108044}{nucl-th/0108044}].

\bibitem{Haberzettl:2011zr}
 H.~Haberzettl, F.~Huang, and K.~Nakayama,
 \emph{Dressing the electromagnetic nucleon current},
 Phys.\ Rev.\ C {\bf 83}, 065502 (2011)
 [\href{https://arxiv.org/abs/1103.2065}{arXiv:1103.2065 [nucl-th]}].

\bibitem{Bernauer:2010zga}
J.~C.~Bernauer, 
\emph{Measurement of the elastic electron-proton cross section
                        and separation of the electric and magnetic form factor in
                        the Q$^{2}$ range from 0.004 to 1 (GeV/c)$^{2}$},
PhD thesis at Johannes Gutenberg Universit\"at Mainz, Institut f\"ur Kernphysik (2010) 
[\href{http://wwwa1.kph.uni-mainz.de/A1/publications/doctor/bernauer.pdf}{http://wwwa1.kph.uni-mainz.de/A1/publications/doctor/bernauer.pdf}].

\bibitem{Higinbotham:2015rja}
D.~W.~Higinbotham, A.~A.~Kabir, V.~Lin, D.~Meekins, B.~Norum, and B.~Sawatzky, 
\emph{Proton radius from electron scattering data}, 
Phys.\ Rev.\ C {\bf 93}, 055207 (2016) 
[\href{https://arxiv.org/abs/1510.01293}{arXiv:1510.01293 [nucl-ex]}].

\bibitem{Frink:2006hx}
 M.~Frink and Ulf-G.~Mei{\ss}ner,
 \emph{On the chiral effective meson-baryon Lagrangian at third
                        order},
 Eur.\ Phys.\ J.\ A29, 255 (2006)
 [\href{https://arxiv.org/abs/hep-ph/0609256}{hep-ph/0609256}].

\bibitem{Oller:2006yh}
J.~A.~Oller, M.~Verbeni, and J.~Prades,
 \emph{Meson-baryon effective chiral lagrangians to O(q**3)},
 JHEP 09, 079 (2006)
 [\href{https://arxiv.org/abs/hep-ph/0608204}{hep-ph/0608204}].

\bibitem{Oller:2007qd}
J.~A.~Oller, M.~Verbeni, and J.~Prades,
 \emph{Meson-Baryon Effective Chiral Lagrangian at O(q**3)
                        Revisited},
 (2007)
 [\href{https://arxiv.org/abs/hep-ph/0701096}{hep-ph/0701096}].

\bibitem{Geng:2009ys}
L.~S.~Geng, J.~Martin Camalich, and M.~J.~Vicente Vacas, 
\emph{Electromagnetic structure of the lowest-lying decuplet resonances in covariant chiral perturbation theory}, 
Phys.\ Rev.\ D {\bf 80}, 034027 (2009) 
[\href{https://arxiv.org/abs/0907.0631}{arXiv:0907.0631 [hep-ph]}].

\bibitem{Hemmert:1996xg} 
T.~R.~Hemmert, B.~R.~Holstein, and J.~Kambor, 
\emph{Systematic 1/M expansion for spin 3/2 particles in baryon chiral perturbation theory}, 
Phys.\ Lett.\ B {\bf 395}, 89 (1997) 
[\href{https://arxiv.org/abs/hep-ph/9606456}{hep-ph/9606456}].

\bibitem{Hemmert:1997ye} 
T.~R.~Hemmert, B.~R.~Holstein, and J.~Kambor, 
\emph{Chiral Lagrangians and delta(1232) interactions: Formalism}, 
J.\ Phys.\ G {\bf 24}, 1831 (1998) 
[\href{https://arxiv.org/abs/hep-ph/9712496}{hep-ph/9712496}].
  
 \bibitem{Unal:2015hea}
 Y.~Unal, A.~Kucukarslan, and S.~Scherer,
 \emph{Interaction of the vector-meson octet with the baryon
                        octet in effective field theory},
 Phys.\ Rev.\ C {\bf 92}, 055208 (2015)
 [\href{https://arxiv.org/abs/1510.02205}{arXiv:1510.02205 [nucl-th]}].

\bibitem{Agashe:2014kda}
 K.~A.~Olive et al. [Particle Data Group],
 \emph{Review of Particle Physics},
 Chin.\ Phys.\ C {\bf 38}, 090001 (2014)
 [\href{http://iopscience.iop.org/article/10.1088/1674-1137/38/9/090001/meta;jsessionid=093DA1F58D861922D568EC99DCE6421C.ip-10-40-1-105}{DOI:10.1088/1674-1137/38/9/090001}].

\bibitem{Dover:1985ba}
 C.~B.~Dover and A.~Gal,
 \emph{HYPERON NUCLEUS POTENTIALS},
 Prog.\ Part.\ Nucl.\ Phys.\ {\bf 12}, 171 (1985)
 [\href{http://www.sciencedirect.com/science/article/pii/0146641084900048?via\%3Dihub}{DOI:10.1016/0146-6410(84)90004-8}].\end{thebibliography}
\end{document}